\documentclass{tcibook}
\usepackage{fancyhea}
\usepackage{work}
\usepackage{bm}       
\usepackage{graphicx}
\usepackage{hyperref}      
\usepackage[usenames,dvipsnames]{color} 
\usepackage{chngcntr}
\counterwithout{figure}{chapter}

\newcommand{\nc}{\newcommand}  

\def\beq{\begin{equation}}
\def\eeq#1{\label{#1}\end{equation}}
\def\eeqn{\end{equation}}

\newenvironment{Eqnarray}%
   {\arraycolsep 0.14em\begin{eqnarray}}{\end{eqnarray}}
\def\beqa{\begin{Eqnarray}}
\def\eeqa#1{\label{#1}\end{Eqnarray}}
\def\eeqan{\end{Eqnarray}}

\nc{\ra}{\rightarrow}  
\nc{\slsh}{\slash\hspace*{-0.22cm}}
\def\Re{{\cal R \mskip-4mu \lower.1ex \hbox{\it e}\,}}
\def\Im{{\cal I \mskip-5mu \lower.1ex \hbox{\it m}\,}}

\nc{\vev}[1]{ \left\langle {#1} \right\rangle }
\nc{\bra}[1]{ \langle {#1} | }
\nc{\ket}[1]{ | {#1} \rangle }
\nc{\fb}{\,{\rm fb}^{-1}}
\nc{\ev}{{\rm eV}}
\nc{\kev}{{\rm keV}}
\nc{\Mev}{{\rm MeV}}
\nc{\gev}{{\rm GeV}}
\nc{\tev}{{\rm TeV}}
\nc{\mev}{{\rm MeV}}

\def\del{\partial}
\def\Dslash{\not{\hbox{\kern-4pt $D$}}}
\def\dslash{\not{\hbox{\kern-2pt $\del$}}}
\def\pslash{\not{\hbox{\kern-2pt $p$}}}
\def\ETmiss{ \not{\hbox{\kern-4pt $E$}}_T }

\def\msb{{\bar{\ssstyle M \kern -1pt S}}}

\newcommand{\cmbexp}{{CMB-S4}}

\begin{document}

\def\bibname{References}

\bibliographystyle{utphys}  

\raggedbottom

\pagenumbering{roman}

\parindent=0pt
\parskip=8pt
\setlength{\evensidemargin}{0pt}
\setlength{\oddsidemargin}{0pt}
\setlength{\marginparsep}{0.0in}
\setlength{\marginparwidth}{0.0in}
\marginparpush=0pt


\pagenumbering{arabic}

\renewcommand{\chapname}{chap:intro_}
\renewcommand{\chapterdir}{.}
\renewcommand{\arraystretch}{1.25}
\addtolength{\arraycolsep}{-3pt}



 
\chapter*{Neutrino Physics from the Cosmic Microwave Background and Large Scale Structure}
\renewcommand*\thesection{\arabic{section}}

\def\nnu{N_{\rm eff}}
\def\gtrsim{\raise-.75ex\hbox{$\buildrel>\over\sim$}}
\begin{center}\begin{boldmath}



\begin{center}

\begin{large} {\bf Topical Conveners: K.N.~Abazajian, J.E.~Carlstrom, A.T.~Lee} \end{large}

K.N.~Abazajian$^\ast$,
K.~Arnold,
J.~Austermann, 
B.A.~Benson,
C.~Bischoff,
J.~Bock,
J.R.~Bond, 
J.~Borrill,
E.~Calabrese, 
J.E.~Carlstrom,
C.S.~Carvalho, 
C.L.~Chang,
H.C.~Chiang, 
S.~Church,
A.~Cooray,
T.M.~Crawford,
K.S.~Dawson,
S.~Das,
M.J.~Devlin,
M.~Dobbs, 
S.~Dodelson,
O.~Dor\'e, 
J.~Dunkley, 
J.~Errard,
A.~Fraisse,
J.~Gallicchio, 
N.W.~Halverson,
S.~Hanany,
S.R.~Hildebrandt,
A.~Hincks, 
R.~Hlozek, 
G.~Holder, 
W.L.~Holzapfel,
K.~Honscheid,
W.~Hu, 
J.~Hubmayr, 
K.~Irwin, 
W.C.~Jones, 
M.~Kamionkowski,
B.~Keating,
R.~Keisler,
L.~Knox,
E.~Komatsu, 
J.~Kovac,
C.-L.~Kuo,
C.~Lawrence,
A.T.~Lee,
E.~Leitch, 
E.~Linder,
P.~Lubin,
J.~McMahon, 
A.~Miller,
L.~Newburgh, 
M.D.~Niemack,
H.~Nguyen,
H.T.~Nguyen,
L.~Page, 
C.~Pryke,
C.L.~Reichardt,
J.E.~Ruhl, 
N.~Sehgal, 
U.~Seljak,
J.~Sievers,
E.~Silverstein,
A.~Slosar,
K.M.~Smith, 
D.~Spergel, 
S.T.~Staggs, 
A.~Stark,
R.~Stompor,
A.G.~Vieregg,
G.~Wang, 
S.~Watson,
E.J.~Wollack,
W.L.K.~Wu,
K.W.~Yoon,
and O.~Zahn
\end{center}

\noindent $^\ast$Corresponding author. Phone: +1 (949) 824-0368, Fax: +1
(949) 824-2174, email: kevork@uci.edu


\end{boldmath}\end{center}

\section*{Executive Summary}

The cosmological background of neutrinos thermally produced in the big
bang has been definitively (albeit indirectly) detected. Measurements
of the cosmic microwave background (CMB) alone have led to a
constraint on the effective number of neutrino species of $N_{\rm
  eff}=3.36 \pm 0.34$~\cite{Ade:2013zuv}, a value $10\sigma$ away from
zero and consistent with expectations.  Experiments planned and
underway are prepared to study this background in detail via its
influence on distance-redshift relations and the growth of structure.
The program for the next decade described in this document, including
upcoming spectroscopic surveys eBOSS and DESI and a new Stage-IV CMB
polarization experiment \cmbexp, will achieve $\sigma(\sum m_\nu) =
16\rm\ meV$ and $\sigma(\nnu) = 0.020$.  Such a mass measurement will
produce a high significance detection of non-zero $\sum m_\nu$, whose
lower bound derived from atmospheric and solar neutrino oscillation
data is about 58 meV.  If neutrinos have a normal mass hierarchy,
this measurement will definitively rule out the inverted neutrino mass
hierarchy, shedding light on one of the most puzzling aspects of the
Standard Model of particle physics -- the origin of
mass. 

This precise a measurement of $N_{\rm eff}$ will allow for a precision test of
the standard cosmological model prediction that $N_{\rm eff} = 3.046$.  The
difference from three is due to the small amount of entropy from
electron/positron annihilation that gets transferred to the neutrinos;
$N_{\rm eff}$ is by design equal to three in the idealized case that all
of this entropy is transferred to photons.  Finding $\nnu$
consistent with 3.046 would demonstrate that we understand very well
the thermal conditions in the universe at just one second after the
big bang.  On the other hand, finding $\nnu$ significantly different
from 3.046 would be a signature of new physics.  Possibilities include
sterile neutrinos produced via oscillations in the early universe, a
large matter/anti-matter asymmetry in the neutrino sector, or new
particles that interact even more weakly than neutrinos but that were
in thermal and chemical equilibrium deep in the radiation-dominated
era (e.g.,~\cite{Weinberg:2013kea}).

The effects of neutrino properties on cosmology are precisely
predicted from theory, appear in many observables in diverse ways, and
are clearly observationally distinguishable from the effects of other
cosmological parameters.  For example, increasing $N_{\rm eff}$ not
only suppresses small-scale fluctuation power in the CMB, but also
leads to detectable changes in the temporal phase of the acoustic
oscillations, e.g.,~\cite{Hou:2012xq}.  Increasing $\sum m_\nu$ in a
manner that is consistent with the measured CMB power spectrum leads
to very specific redshift- and scale-dependent changes to the power
spectrum of both matter and the galaxy
distribution~\cite{Lesgourgues:2006nd}.  These signatures will provide
important consistency tests.

Much of the sensitivity to $\sum m_\nu$ will come from measurements of
the gravitational lensing of the CMB, measurements of the baryon
acoustic oscillation (BAO) features and broadband power spectrum, and
measurements of weak gravitational lensing of galaxies. Different
combinations of these probes have completely independent systematic
errors.

 Significant progress has been made recently in these areas,
 with the first projected mass reconstructions from CMB lensing
 \cite{Das:2011ak,vanEngelen:2012va,Das:2013zf,Holder:2013hqu,Ade:2013tyw,Geach:2013zwa},
 detection of the CMB lensing $B$-mode polarization
 \cite{Hanson:2013daa}, and percent level measurements of the distance
 scale from BAO measurements
 \cite{Blake:2012pj,Anderson:2012sa,Busca:2012bu,Slosar:2013fi}.

\section{Introduction}
\label{sec:neutrino-intro}
\subsection{ Motivation}
One of the most remarkable aspects of physical cosmology is that the
study of the largest physical structures in the universe can reveal
the properties of the particles with the smallest known cross section:
neutrinos.  At its root, this cosmological sensitivity to neutrino
properties is due to the fact that the neutrino cosmological number
density is second only to CMB photons. Coupled with their non-zero
mass, the large number of neutrinos leads to an energy density today
at least 25 times larger than the CMB. This high density of free
streaming particles rapidly inhibits the growth of structure at late
times, leading to changes in large scale structure (LSS) that can be
captured in large galaxy surveys or CMB lensing measurements.

The CMB and LSS are sensitive to the total energy density of cosmic
neutrinos today, which is effectively a measure of the sum of the
neutrino masses, since the number density is known (up to
uncertainties in the CMB temperature and $\nnu$). These Cosmic
Frontier experiments are then exactly complementary to terrestrial
laboratory experiments on the Intensity Frontier that measure the
differences of mass squared and CP-violation in the neutrino sector.

Short baseline neutrino oscillation results hint at a richer neutrino
sector with more than three active neutrinos participating in flavor
oscillations, i.e. with one or more sterile
flavors~\cite{Aguilar:2001ty,AguilarArevalo:2010wv,Kopp:2011qd,Giunti:2011gz}. The
reactor neutrino anomaly may also indicate the presence of light
sterile neutrinos~\cite{Mention:2011rk,Huber:2011wv}.  If they do
indeed exist, these extra degrees of freedom would affect the cosmic
neutrino background, and the relativistic energy density in the early
universe.  As we will discuss below, future CMB-S4 and LSS experiments
in the Cosmic Frontier have the sensitivity to make a high-confidence
detection of more than three standard neutrinos or new relativistic
degrees of freedom in a new particle sector behaving like neutrinos.

A global fit to solar and atmospheric neutrino flavor oscillations
implies a difference in the squares of masses
$m_2^2-m_1^2=7.54^{+0.26}_{-0.22} \times 10^{-5}$ eV$^2$ and $m_3^2-
(m_1^2+m_2^2)/2= \pm 2.43^{+0.06}_{-0.10} \times 10^{-3}$ eV$^2$
\cite{Fogli:2012ua}.  For the minimal case of a normal hierarchy,
where $m_1 < m_2 \ll m_3$, the smallest possible value for the sum of
the masses occurs when $m_1 = 0$, $m_2 =
8.68^{+0.15}_{-0.13}\rm\ meV$, and $m_3 = 49.7^{+0.8}_{-1.0}
\rm\ meV$.  So, in the normal hierarchy, the minimum sum of the masses
is $\sum m_\nu = 58.4^{+1.2}_{-0.8}\rm\ meV$.  In the case of a
so-called inverted hierarchy, where $m_1\simeq m_2 \gg m_3$, the
minimum sum of the masses must be greater than 100 meV.  For the
degenerate neutrino mass case where $m_1 \simeq m_2 \simeq m_3$, the
sum of neutrino masses is at least approximately 150 meV.  As we will
discuss below, future CMB-S4 and LSS experiments in the Cosmic
Frontier have projected constraints to detect the minimum mass scale
of 58 meV at $\sim$4$\sigma$ confidence, a ground-breaking result.

\begin{figure}[t!]
\begin{center}
\includegraphics[width=0.8\hsize]{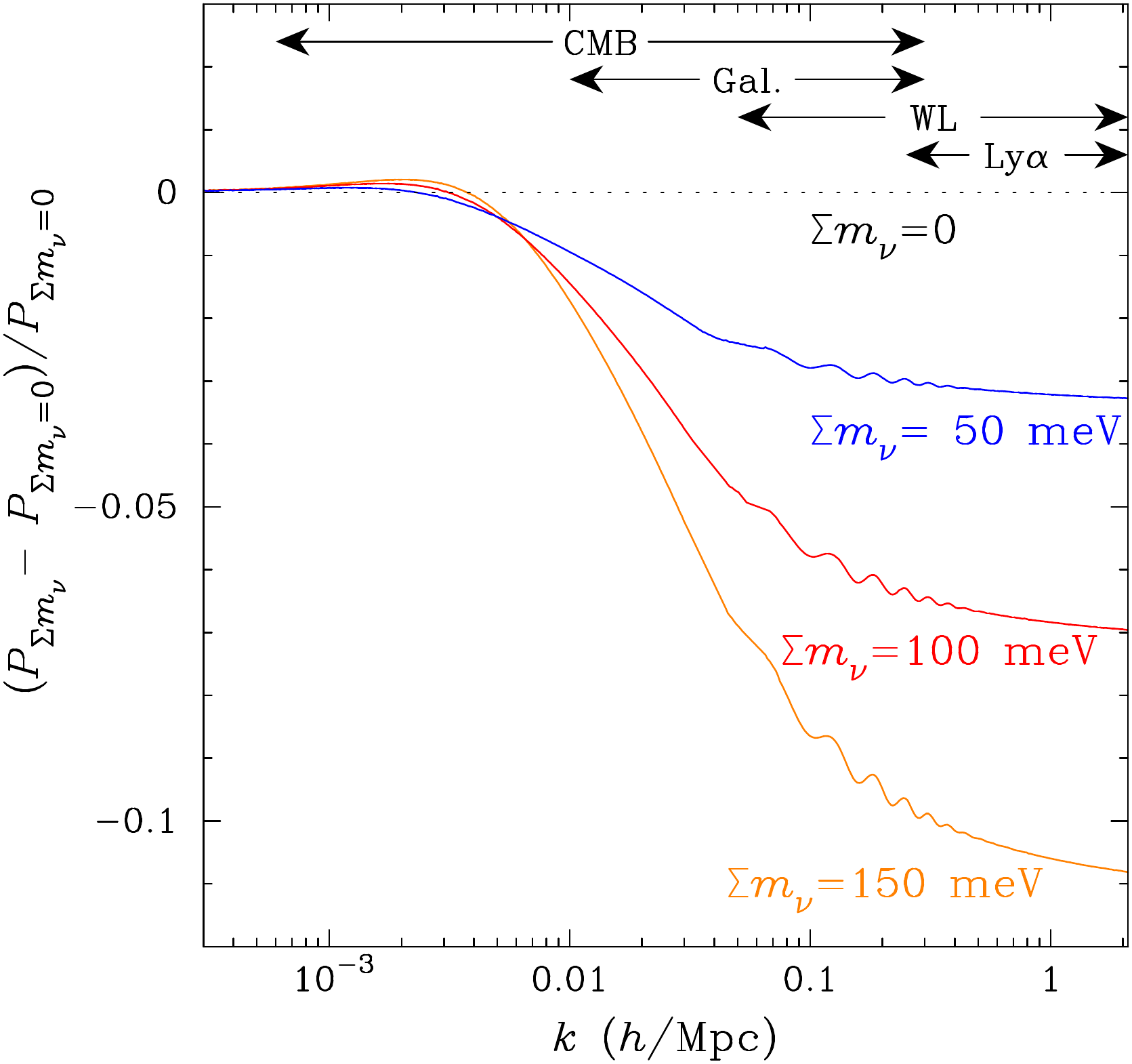}
\caption{ Fractional change in the matter density power spectrum as a
  function of comoving wavenumber $k$ for different values of $\sum
  m_\nu$.  Neutrino mass suppresses the power spectrum due to free
  streaming below the matter-radiation equality scale.  The shape of
  the suppression is highly characteristic and precision observations
  over a range of scales can measure the sum of neutrino masses (here
  assumed all to be in a single mass eigenstate).  Also shown are the
  approximate ranges of experimental sensitivity in the power spectrum
  for representative probes: the cosmic microwave background (CMB),
  galaxy surveys (Gal.), weak lensing of galaxies (WL), and the
  Lyman-alpha forest (Ly$\alpha$). The CMB lensing power spectrum
  involves (an integral over) this same power spectrum, and so is also
  sensitive to neutrino mass.}
\label{fig:matterpower}
\end{center}
\end{figure}

\subsection{The Effective Number of Neutrino Species, $\nnu$}

\label{physicsofnnucosmo}

Relic primordial neutrinos leave a distinct signature in the CMB.
Assuming that all entropy produced by electron-position annihilation
is transferred to photons, it can be shown, by considering the
available degrees of freedom, that the temperature of the photon gas
is a factor of $(11/4)^{1/3}$ higher than the temperature of the
neutrino gas. The total cosmological number density of neutrinos (and
antineutrinos) is given by

\begin{equation}
n_\nu = \nnu \left(\frac{3}{4}\right)\left(\frac{4}{11}\right) n_\gamma,
\label{eq:nunumberdens}
\end{equation} 
where $n_\gamma$ is the density of the CMB photons, $\nnu$ is the {\it
  effective} number of neutrino species in the universe, and the
factor of 3/4 comes from the difference between Fermi-Dirac and
Bose-Einstein statistics. The relativistic energy density in the early
universe including neutrinos is given by
\begin{equation}
  \rho_R = \left(1 + \nnu \frac{7}{8}\left(\frac{4}{11}\right)^{4/3}
  \right) \rho_\gamma,
\end{equation}
where the factor $7/8$ accounts for the fact that neutrinos are
fermions.

$\nnu$, as defined above, would be exactly equal to three if neutrinos
instantaneously decoupled from the primordial plasma.  However, $\nnu$
differs from the integer three, due to known properties of neutrinos
in the early universe.  First, accurate calculations show that
neutrinos are still interacting with the primordial plasma when the
process of electron-positron annihilation begins. Second, the energy
dependence of neutrino interactions causes the high-temperature tail
of the Fermi-Dirac distribution to interact more strongly, leading to
an energy-dependent distortion in the energy spectrum of neutrino gas.
These effects conspire to raise $\nnu$ to $\nnu=3.046$
\cite{Mangano:2005cc}. Once the decoupling has completed, the
relativistic neutrino gas redshifts while keeping its distorted energy
spectrum shape\footnote{Note that in the case of massive neutrinos,
  the spectrum will be altered when the neutrinos start becoming
  nonrelativistic and the evolution must be calculated with general
  relativistic kinetic theory~\cite{BernsteinBook1988}, which is
  handled properly by modern Boltzmann solvers.}. Therefore, in the
absence of non-gravitational interactions, the relativistic neutrino
gas can be described by the modified effective number of neutrino
species, $\nnu$.  A high-confidence level determination of a neutrino
energy density consistent with 3.046 away from the integer value would
be a striking confirmation of standard cosmology, its thermal history,
and convincingly show that the measured energy density is composed of
Standard Model neutrinos.

Beyond Standard Model physics can also change $\nnu$. A particular
example are additional ``sterile'' neutrino flavors, but in general
any relativistic degrees of freedom that interact solely
gravitationally (also dubbed \emph{dark radiation}) will raise $\nnu$.
A high-confidence deviation from the standard value of $\nnu=3.046$
would be a strong indication of new neutrino physics, new particles,
or an invalidity in the theoretical assumptions going into the
standard cosmological theory.

The number of neutrino species primarily affects the CMB by altering
the photon diffusion (Silk damping) scale, $r_d$, relative to the
sound horizon, $r_s$.  The sound horizon sets the location of the
acoustic peaks while photon diffusion suppresses power at small
angular scales.  CMB measurements are sensitive to both as angles
projected over the distance to last scattering: $\theta_s = r_s/D_A$
and $\theta_d = r_d/D_A.$ Sensitivity of these observables to the
number of neutrino species comes from how the energy density of the
neutrinos increases the expansion rate.  The sound horizon scales as
$1/H$, while the diffusion scale, resulting from a stochastic process,
scales instead as $\sqrt{1/H}$.  Thus, the CMB can measure the
expansion rate in the early universe, and therefore the relativistic
energy density, by measuring the ratio $\theta_d/\theta_s = r_d/r_s
\propto \sqrt{H}$ (see \cite{Hou:2013a} and references therein).

Due to this effect one can infer $\nnu=3.36 \pm
0.34$~\cite{Ade:2013zuv} from the combination of temperature data from
Planck, WMAP polarization data \cite{Bennett:2012zja}, and high-$\ell$
CMB measurements (South Pole Telescope (SPT) \cite{Story:2012wx} and
Atacama Cosmology Telescope (ACT) \cite{Das:2013zf}).  LSS
measurements improve these constraints by breaking degeneracies with
other cosmological parameters.  In \S\ref{sec:forecast}, we will give
the constraints from combining future CMB and LSS measurements.

\subsection{The Sum of the Neutrino Masses, $\sum m_\nu$}
\label{physicsofnucosmo}

The physical effects of massive neutrinos on the CMB and LSS have been
studied for nearly two decades (e.g., \cite{Jungman:1995bz}).  When
the detailed effects of the different components of matter in the
universe on linear LSS were first precisely calculated
(e.g.,~\cite{Eisenstein:1997jh}), it became clear that the impact of
massive neutrinos on LSS was quite large, making LSS a sensitive probe
of massive neutrinos~\cite{Hu:1997mj}.  The physical effect of massive
neutrinos on LSS is determined by the fact that massive neutrinos
behave as radiation-like particles in the early universe, and as
matter-like particles in the late universe
\cite{Dodelson:1995es,Hu:2001bc,Hou:2012xq}.  Since the number density
of neutrinos is comparable to that of photons
(Eq.~\ref{eq:nunumberdens}), they will contribute considerably to the
relativistic energy density when their energies are relativistic.  If
the neutrinos are to affect matter clustering in a nontrivial amount,
then they must have contributed a nontrivial amount to the total
matter density.  Because of the neutrinos' high cosmological number
density, a small neutrino mass lets them contribute to the critical
density as
\begin{equation}
\Omega_\nu h^2 \simeq \frac{\sum m_\nu}{93\rm\ eV}.
\label{eq:omeganu}
\end{equation}
Therefore, even the minimum summed neutrino mass of 58 meV contributes
to a matter fraction $f_\nu~\equiv~\Omega_\nu/\Omega_m$ of about
0.4\%.  

In broad terms, the neutrinos cool down by the expansion of the
universe and transition from being ultra-relativistic gas behaving as
radiation to being cold gas behaving as dark matter at a redshift
around $z_{\rm nr}\sim 2000 m_\nu/1{\rm eV}$. Before they are
non-relativistic, they free-stream out of over-dense regions, erasing
the structure on small scales. This free-streaming establishes a
preferred scale, which very roughly corresponds to the horizon scale
at time of transition to the non-relativistic matter. In models with
massive neutrinos, power appears suppressed on scales smaller than the
free-streaming scale when normalizing the matter power spectrum at
large scales. This scale numerically turns out to be close to the peak
in the matter power spectrum, a feature produced because of the
transition from cosmological radiation to matter domination when
perturbations cross the cosmological horizon. This effect is shown in
Figure~\ref{fig:matterpower}.  Measurements of this feature in
cosmological LSS can be made in two ways: first through the overall
change in shape of the power spectrum, or second in a relative
amplitude measurement, where for example the CMB precisely measures
the amplitude at large scales and a LSS probe gives a measure at small
scales.

Among the most stringent current combined CMB plus LSS cosmological
constraints on $\sum m_\nu$ result from the combination of Planck CMB
data, WMAP 9-year CMB polarization data \cite{Bennett:2012zja}, and a
measure of baryon acoustic oscillations (BAO) from the Baryon
Oscillation Spectroscopic Survey (BOSS) \cite{Dawson:2012va}, Sloan
Digital Sky Survey (SDSS), Wigglez LSS data, and the 6dF galaxy
redshift survey.  As reported in \cite{Ade:2013zuv}, these combined
probes produce an upper limit $\sum m_\nu < 0.23\rm\ eV$(95\% C.L.).
A more aggressive use of galaxy clustering into smaller scales and the
nonlinear clustering regime can lead to more stringent constraints
(e.g.,~\cite{Zhao:2012xw,Riemer-Sorensen:2013jsa}).

Sensitivity to effects of neutrinos on cosmological gravitational
perturbation evolution and the matter power spectrum are measured in
several ways:

\noindent {\bf Probing $\sum m_\nu$ with CMB temperature and
  polarization}--- The CMB constrains the neutrino mass through its
effect on structure growth in two primary ways: 1) the early
Integrated Sachs Wolfe (ISW) effect, and 2) gravitational lensing of
the CMB by LSS.  A significant fraction of the power in the CMB on a
degree angular scale is from the early ISW effect.  This measurement
has been used to infer $\sum m_\nu < 0.66\rm\ eV$ (95\%
C.L.)~\cite{Ade:2013zuv}. Only marginal improvement is possible
with that technique as the constraint is already limited by cosmic
variance due to its reliance on the large angular scale power in the
CMB.

Gravitational lensing of the CMB provides a clean and direct
measurement of the matter power spectrum on scales where effects of
neutrino mass manifest.  Recently, ACT, SPT, and Planck have used high
order statistics in temperature anisotropy to produce lensing maps.
Polarization surveys will represent a large improvement over what can
be done with temperature.  Gravitational lensing of primordial
$E$~modes by intervening matter distributions produces a $B$-mode
polarization signal \cite{Zaldarriaga:1998ar}, which has recently been
detected \cite{Hanson:2013daa}.  This lensing-induced $B$-mode
polarization corresponds to an RMS of $\sim 5\;\mu$K-arcmin with a
characteristic angular scale is a few arcminutes. Detailed, high
signal-to-noise measurements of arcminute polarization can therefore
be used to reconstruct the lensing potential.  This effect has already
been used to achieve modest improvements in the CMB-based neutrino
mass constraints \cite{Story:2012wx,Das:2013zf,Ade:2013zuv}, which we
expect to improve by nearly an order of magnitude with a Stage-IV CMB
experiment (see \S\ref{sec:forecast}).

\noindent {\bf Probing $\sum m_\nu$ with baryonic tracers of the large
  scale clustering of matter}--- Galaxy surveys, 21~cm surveys, and
the Lyman-$\alpha$ forest are measurements of tracers of the
underlying matter clustering in LSS.  It can be shown that the large
scale fluctuations in the tracer will follow those of the dark matter
with a scale-independent constant of proportionality as long as
processes that determine the local number density of galaxies and the
emission of 21~cm radiation among other observables are local
functions of the dark matter density~\cite{Narayanan:1998wd}. On
smaller scales the biasing relation can become more complicated.  In
the case of 21~cm and the Lyman-$\alpha$ forest, neutral hydrogen gas
is the tracer, which, to zeroth order, should follow the underlying
dark matter distribution. In the case of the Lyman-$\alpha$ forest,
the gas is seen via absorption features toward a distant quasar, and
therefore the measure is a one-dimensional tracer. In the case of
spectroscopic galaxy surveys, the tracer is that of galaxies, which
have been found to have scale-independent bias at large scales, both in
observations~\cite{Tegmark:2003uf} and in simulations,
e.g. Ref.~\cite{Narayanan:1998wd}.

Additional information comes from the fact that we are observing
tracers in redshift-space rather than real-space, which distorts the
observed two-point function in the direction along the line of
sight. Since galaxies cannot experience velocity bias on very large
scales, this distortion allows one to break the degeneracy between
bias and the amplitude of dark matter
fluctuations~\cite{McDonald:2008sh}. This further improves our ability
to constrain neutrino properties by comparing the small-scale power
amplitude measured by galaxy clustering and the large scale power
amplitude determined by the CMB. In Fisher matrix projections
described below these constraints emerge naturally by considering the
anisotropic measurements of the galaxy power spectrum. Further details
of this method are given in Ref.~\cite{Font-Ribera:2013rwa}.

More details are given in \S\ref{galaxyforecast} \&
\ref{otherlssforecast}.

\noindent{\bf Probing $\sum m_\nu$ with weak lensing}--- Deep, high
angular resolution observations of the sky can reveal gravitational
weak lensing of background galaxies by foreground LSS, sometimes
dubbed cosmic shear.  The prospect of the statistical angular and
tomographic correlations in the mean weak lensing signal for being a
precision probe of LSS has been studied for some
time~\cite{Kaiser:1991qi,Jain:1996st,Hu:1999ek}.  The use of the weak
lensing signature to infer the neutrino masses was first discussed via
angular correlation measures in the deep nonlinear clustering
regime~\cite{Cooray:1999rv} and in tomographic correlations in the
linear regime~\cite{Abazajian:2002ck}.  The original conservative
linear-regime clustering forecast of Ref.~\cite{Abazajian:2002ck}
determined a sensitivity of $\sum m_\nu < 300\rm\ meV$ for a 4000
square degree weak lensing survey, with a depth of one hundred
galaxies per square arcminute, potentially achievable by space-based
surveys, in combination with cosmological parameter constraints from
Planck CMB observations.

\section{ Forecast sensitivity to $\nnu$ and $\sum m_\nu$} 
\label{sec:forecast}

In this section, we present forecast sensitivity to cosmic neutrinos
from future CMB and LSS experiments.  Various complementary probes of
neutrino mass are surveyed.  However, we highlight a few methods
considered to be least sensitive to systematic effects: an
arcminute-scale CMB polarization survey from the Stage-IV CMB
experiment \cmbexp\ (described below and in the Appendix), galaxy
clustering, and cosmic shear.  The exact forecast of uncertainty
depends on the cosmology and priors assumed.  Our main objective in
this Community Planning Study is to provide reliable comparisons
between current and expected limits, and to emphasize the improvements
and degeneracy-breaking capabilities of complementary probes.  To meet
this goal, efforts are made to consistently and systematically
calculate these numbers using multiple extensively checked Fisher
matrix pipelines, even though some forecast numbers disagree slightly
with what is in the literature. 

We warn the reader of the usual caveats associated with Fisher matrix
formalism when forecasting the reach of future experiments. This
formalism makes a number of simplifying assumptions including that
likelihoods and posteriors are Gaussian.  Some of these methods can be
improved by, for example, forecasting through full likelihood
evaluations using Markov Chain Monte Carlo methods and mock future
data (e.g.,~\cite{Wolz:2012sr, Khedekar:2012sh}).  Moreover, for a
lower measured mass, the posterior will have to take into account the
physical limit $\sum m_\nu>0$, which will thus relax constraints with
respect to the Fisher forecasts.  In our forecasts, we use a single
massive neutrino and two massless neutrinos for simplicity in not
modeling the neutrino mass spectrum as well. When varying $N_{\rm
  eff}$, we vary the massless neutrino density. Since the forecast
sensitivity will not probe the exact neutrino mass spectrum directly,
this method suffices \cite{Laureijs:2011gra}.

We also note that we assume a
maximum usable multipole of $\ell_{\rm max}=5,000$ in the
analysis. (Note, here we refer to $\ell$ as CMB multipoles and
$\ell^{rm WL}$ for weak lensing multipoles, below.) We find that the
cosmological parameter constraints are well converged for an
$\ell_{\rm max}$ of $4,000$, and that the range from $\ell_{\rm
  max}=3,000-4,000$ contains a small amount of (mostly polarization)
information. It should be prudent to use this multipole range in our
forecasts, since estimates of polarized foregrounds show these to be
much smaller than the lensed primordial CMB, relatively speaking,
unlike the foregrounds for temperature fluctuations (see e.g.,
\cite{Smith09}).

\begin{figure}[t]
\begin{center}
\includegraphics[width=0.8\hsize]{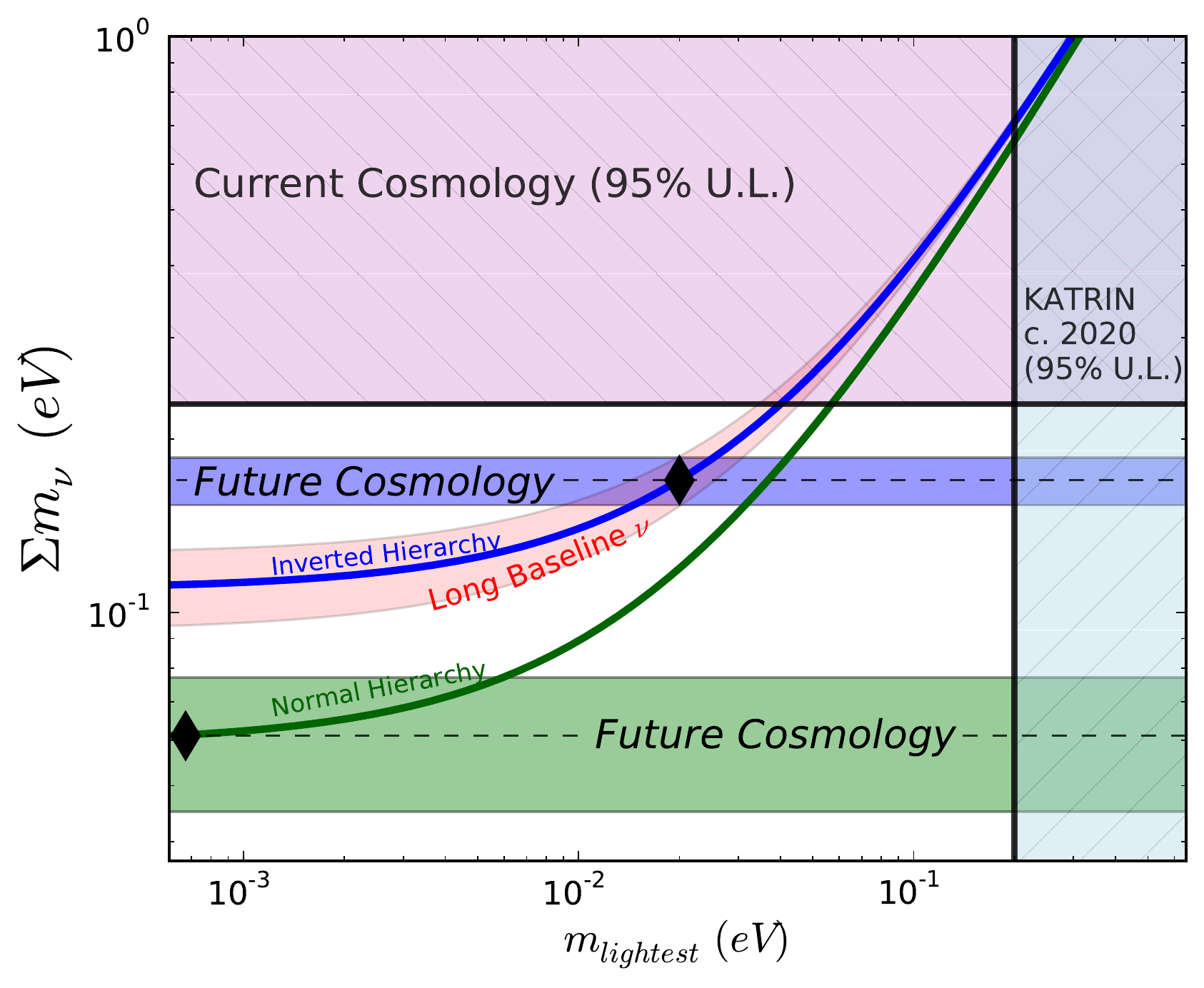}
\caption{Shown are the current constraints and forecast sensitivity of
  cosmology to the neutrino mass in relation to the neutrino mass
  hierarchy.  In the case of an ``inverted hierarchy,'' with an
  example case marked as a diamond in the upper curve, future combined
  cosmological constraints would have a very high-significance
  detection, with $1\sigma$ error shown as a blue band.  In the case
  of a normal neutrino mass hierarchy with an example case marked as
  diamond on the lower curve, future cosmology would detect the lowest
  $\sum m_\nu$ at a level of $\sim 4 \sigma$. Also shown is the
  sensitivity from future long baseline neutrino experiments as the
  pink shaded band, which should be sensitive to the neutrino
  hierarchy at least at $3 \sigma$ \cite{Adams:2013ita}.
\label{fig:hierarchy}}
\end{center}
\end{figure}

It is found in this study that for each of the probes, the expected
$1\sigma$ uncertainty on neutrino mass lies in the range of $20-30$
meV.  When multiple probes are combined, for example for DESI+\cmbexp,
the uncertainty reduces to $\sim 15$ meV.  As shown in
Figure~\ref{fig:hierarchy}, such a joint program will unambiguously
detect neutrino mass under both hierarchy scenarios, since $\sum
m_\nu$ is already known to be at least $58$ meV.

\begin{figure}[t!]
\begin{center}
\includegraphics[width=0.7\linewidth]{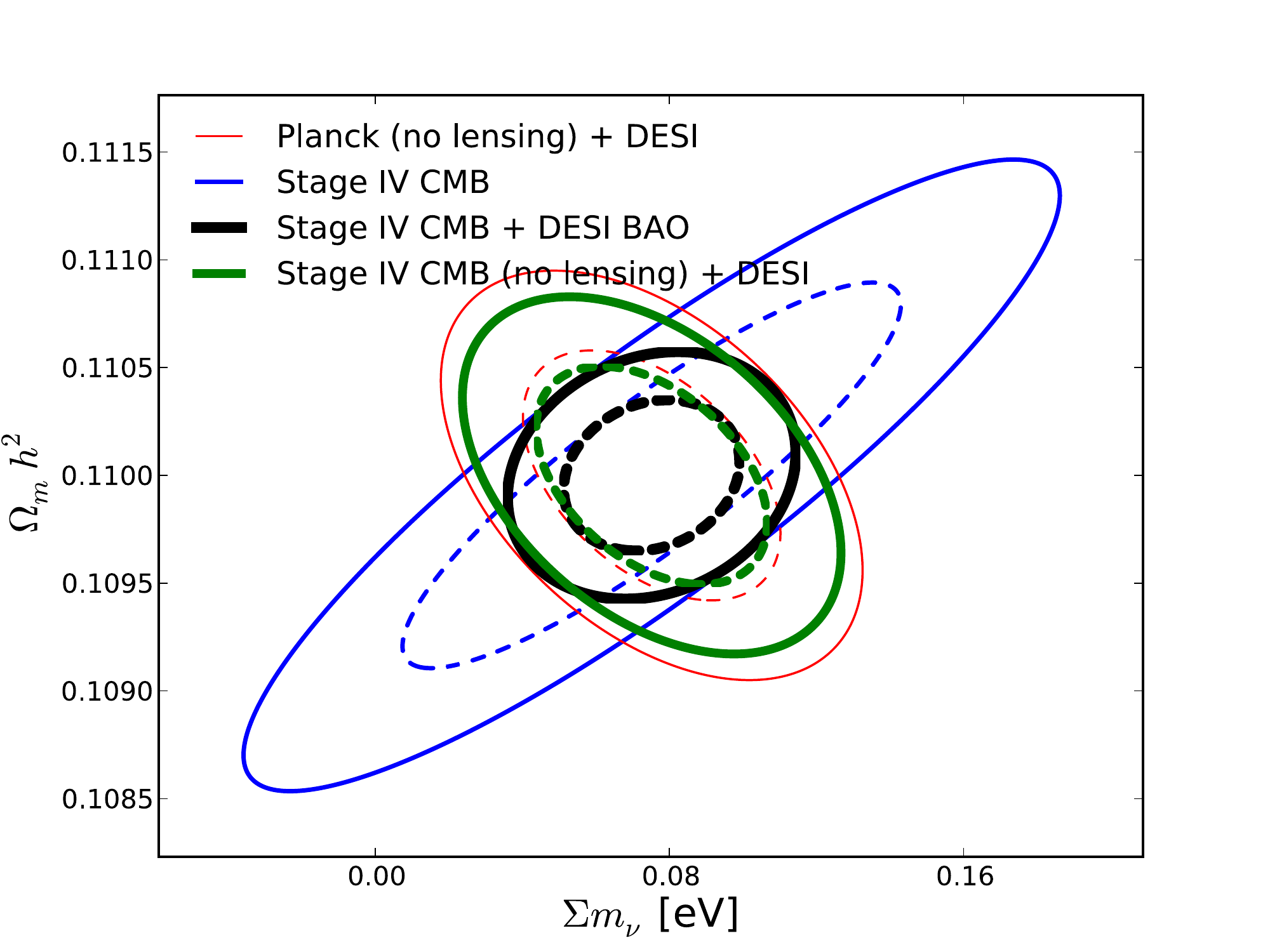}

\caption{Forecasted $1 \sigma$ and $2 \sigma$ constraints in the
  $\Sigma m_\nu - \Omega_m h^2$ plane showing the synergy of an
  experiment like DESI and a Stage-IV CMB lensing experiment.  For
  contrast, a combination of $\it Planck$ data with the lensing
  information removed and DESI are shown in the red contours.  The
  blue contours show the constraint generated by the CMB lensing
  experiment, corresponding to a $24$~meV constraint on massive
  neutrinos. The black contours show the result of adding only DESI
  BAO information, resulting in a 16~meV constraint. This can be
  compared to the case where no Stage-IV CMB lensing information but
  all galaxy clustering information is used, yielding a 24 meV
  constraint. The combination of a Stage-IV CMB experiment and BAO
  information from DESI should allow a robust measurement of the sum
  of the neutrino masses.
\label{fig:Mnu-Omh2}}
\end{center}
\end{figure}

\begin{figure}[ht]
\begin{center}
\includegraphics[width=0.7\linewidth]{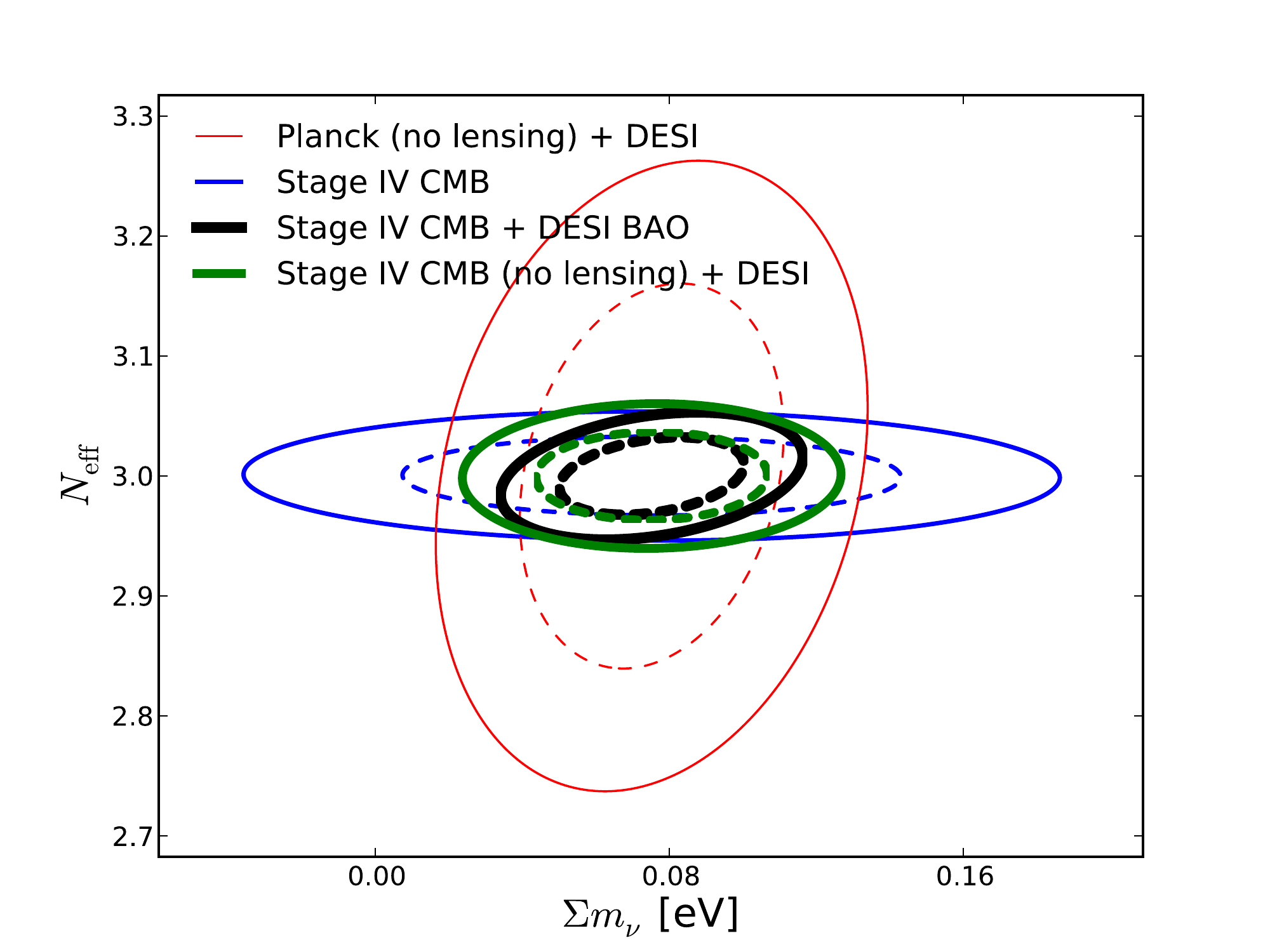}
\caption{The same as Figure \ref{fig:Mnu-Omh2}, but showing forecasts 
  in the $\Sigma m_\nu$ - $N_{\rm eff}$ plane for
  a model including the effective number of neutrino species as a free parameter.
  A Stage-IV CMB experiment will
  not be able to distinguish between the standard model value of
  $N_{\rm eff}=3.046$ and the integer value of $3$ at high
  statistical significance, but it will indicate a preference for one over
  the other at the $\sim 2~\sigma$ level.
\label{fig:Mnu-Neff}}
\end{center}
\end{figure}

Figure~\ref{fig:Mnu-Omh2} shows the interplay between CMB and
galaxy-survey constraints on $\sum m_\nu$.  There is a degeneracy
between neutrino mass and the matter density in the CMB lensing
constraint -- both parameters primarily affect the broadband amplitude
of the lensing effect, with the neutrino mass also introducing a small
scale dependence (see Figure~\ref{fig:matterpower}). A low-redshift
constraint on the acoustic peak scale from, e.g., DESI (see Section
\ref{galaxyforecast} for the survey description), significantly
reduces this degeneracy.  Because CMB lensing and galaxy clustering
constrain the neutrino mass in a similar way by using some of the same
information (see Figure~\ref{fig:matterpower}), we take a conservative
approach when combining CMB and galaxy surveys and only use either
galaxy clustering or CMB lensing information.  When CMB lensing is
included, we omit clustering information in the galaxy survey and use
only the BAO constraint on the acoustic scale at low redshift.  When
CMB lensing information is omitted, we use both BAO and power spectrum
shape information from the galaxy surveys. 

Figure~\ref{fig:Mnu-Neff} shows forecast constraints in the $\sum
m_\nu - N_{\rm eff}$ plane . Note that we do not expect that there is
a physical motivation for these two parameters to be varied at the
same time -- we simply do this because neutrino mass is an unknown
quantity that needs to be marginalized over. The majority of
information in this case comes from precise measurements of the photon
diffusion scale relative to the sound horizon scale as described in
the previous section. Here, the addition of high accuracy $E$-mode
polarization measured to fine angular scales allows these two
quantities to be measured with sufficient precision to decrease the
error bars several-fold with respect to Planck data. The addition of
broadband galaxy power spectra does not help in this case.

This accuracy will not allow us to distinguish between $N_{\rm eff}=3$
and $N_{\rm eff}=3.046$ at more than $2\sigma$. However, we note that
even if the true value of $N_{\rm eff}$ is not $3.046$, it is highly
unlikely to be the unphysical value of a simplified model $N_{\rm
  eff}=3$. We argue that the error on $N_{\rm eff}$ is of the same
order of magnitude as typical corrections stemming from detailed
modeling of the thermodynamical processes in the early universe and
therefore we are sensitive to the non-standard physics that would
produce a signal in $N_{\rm eff}$ order or larger than those
processes.

\begin{table*}
\begin{center}
\caption{Projections for neutrino mass and $N_{\rm eff}$. Projections
  for neutrino masses are assuming a standard value of $N_{\rm
    eff}$. Projections for $N_{\rm eff}$ are for marginalizing over
  neutrino mass for galaxy clustering and CMB lensing. All values are
  forecasts from analyses for this work, except for the projections
  for galaxy weak lensing in the last two rows, which are taken from
  literature and referenced. All errors are 68\% Fisher matrix
  predictions. In the case of two numbers $a/b$, these correspond to
  optimistic/conservative cases with $k_{\rm max}=0.2\ h\ \rm
  Mpc^{-1}$ and $k_{\rm max}=0.1\ h\rm\ Mpc^{-1}$, respectively. All
  values are forecasts from analyses for this work, except for the
  last two rows, which are referenced.  Numbers that we want to
  highlight in this report are in bold.}
\label{tab:tab}
\vspace*{1cm}
\begin{tabular}{lcc}
Dataset  & $\sigma \left(\sum m_\nu\right)$ [meV] & $\sigma \left(N_{\rm eff}\right)$  \\ 
\hline
\\
\textbf{Galaxy Clustering (current CMB):} &&\\
Planck + BOSS BAO   & 100  &  0.18 \\
Planck + BOSS galaxy clustering  & 46/68 & 0.14/0.17 \\
Planck + eBOSS BAO & 97 & 0.18 \\
Planck + eBOSS galaxy clustering & 36/52 & 0.13/0.16 \\
Planck + DESI BAO & 91 & 0.18 \\
Planck + DESI galaxy clustering & \textbf{17/24} & 0.08/0.12\\ 
\\
\textbf{CMB Lensing (current galaxy clustering):} &&\\
Stage-IV CMB &   45 & \textbf{0.021}\\
Stage-IV CMB + BOSS BAO & \textbf{25}&  0.021\\
\\
\textbf{CMB Lensing + Galaxy clustering:}\\
Stage-IV CMB + eBOSS BAO & 23 & 0.021 \\
Stage-IV CMB + DESI BAO & \textbf{16} & 0.020\\
Stage-IV CMB no lensing + DESI galaxy clustering & 15/20 & 0.022/0.024\\
\\
\textbf{Galaxy Weak Lensing:}&&\\
Planck + LSST \cite{Joudaki:2011nw} & 23 & 0.07 \\
Planck + Euclid \cite{Laureijs:2011gra} & 25 & NA$^\dagger$\\
\\
{\footnotesize $^\dagger$Ref.~\cite{Laureijs:2011gra} did not include a forecast for $\nnu$.}\\
   \end{tabular}
   \end{center}
   \end{table*}

\subsection{Lensing of the CMB}

\label{cmbforecast}

In the next decade, ground based CMB experiments will make significant
progress in measuring the effects of gravitational lensing on the CMB
polarization \cite{Kaplinghat:2003bh}.  Current and upcoming CMB
experiments are classified into stages with each stage corresponding
to an order of magnitude increase in sensitivity, or equivalently, the
number of measured modes. Stage-II experiments measure $O$(1000)
optical modes and are now providing the first statistical detections
of lensing $B$-mode polarization \cite{Hanson:2013daa}.  Many of the
expected properties of the lensing signal, such as its amplitude and
statistics, will be measured with real data for the first time in the
Stage-II era. Stage-III experiments will deploy in the latter half of
the decade and measure $O$(10,000) modes. These experiments will be
the first instruments with sufficient sensitivity to make high
signal-to-noise maps of the CMB lensing modes over a few thousands
square degrees of sky. Stage III will correspond to a transition of
the CMB lensing measurement from a statistical detection into
imaging. A Stage-IV experiment observing $O$(100,000) modes is
anticipated to deploy in $\sim$2020. Stage-IV experiments will build
on Stage III to map tens-of-thousands of square degrees of sky to a
depth of $\le 1 \mu \mathrm{K}$ per 1-arcminute pixel with an angular
resolution of $\le 3$ arcminutes. In the Appendix, we describe the
technical program for \cmbexp, a Stage-IV CMB experiment which aims to
deploy in 2020 and operate for several years. One of the most exciting
prospects of a Stage-IV experiment like \cmbexp\ is to measure the
signatures of cosmic neutrinos with high precision.

CMB polarization as a probe of large scale structures has a few unique
advantages.  First of all, CMB lensing is highly complementary to
galaxy surveys, since it probes matter distributions in the linear
regime at higher redshift ($z\sim 2-4$).  Secondly, because the
unlensed {\em background} is precisely understood
(Gaussian-distributed E-mode polarization at redshift $z=1090$ in the
absence of non-Gaussianities, which are strongly limited in the
primordial CMB~\cite{Ade:2013ydc}), the reconstruction of lensing
potential is absolutely calibrated and free of shape noise.  This
property also enables reconstruction beyond the quadratic order, with
sensitivity only limited by instrumental noise.  Finally, the
systematics associated with CMB lensing originated largely from
well-understood instrumental effects, which tend to decrease with
higher resolution.

\begin{figure}[t!]
\begin{center}
\includegraphics[width=0.8\hsize]{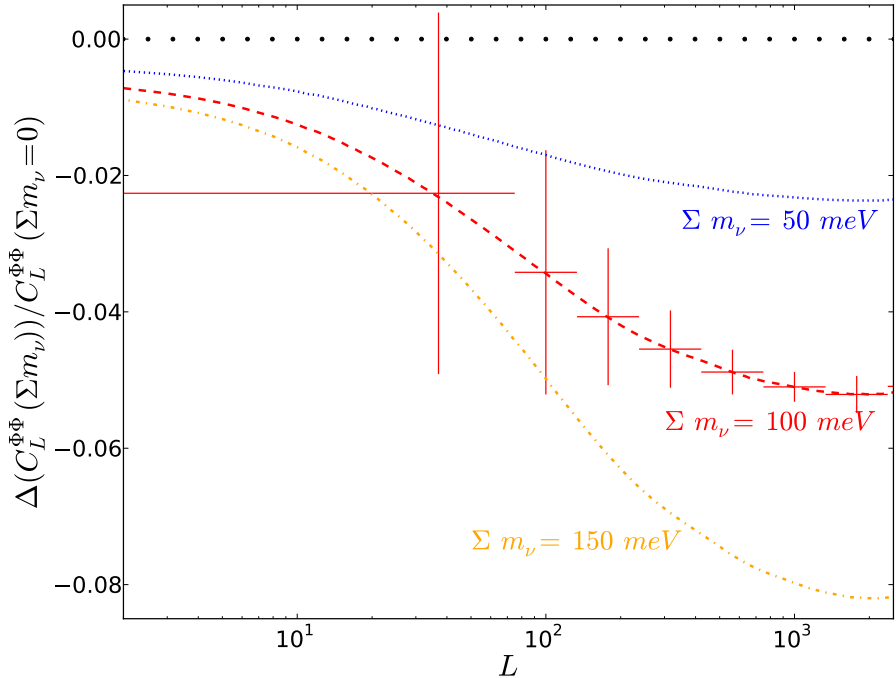}
\caption{ The effect of massive neutrinos on the CMB lensing potential
  power spectrum $C_L^{\Phi \Phi}$. The fractional change in
  $C_L^{\Phi \Phi}$ for a given value of $\sum m_\nu$ is shown
  relative to the case for zero neutrino mass.  Projected constraints
  on $C_L^{\Phi \Phi}$ for a Stage-IV CMB experiment are shown for
  $\sum m_\nu = 100 \ \mathrm{meV}$. Here we have approximated all of
  the neutrino mass to be in one mass eigenstate and fixed the total
  matter density $\Omega_m h^2$ and $H_0$. The $1\sigma$ constraint
  for $\sum m_\nu$ is approximately 45 meV for lensing alone and drops
  to 16~meV when combined with other probes.}
\label{fig:cmblensing}
\end{center}
\end{figure}

Figure \ref{fig:cmblensing} shows the projected constraints on the CMB
lensing potential power spectrum $C_L^{\Phi \Phi}$ for a Stage-IV CMB
experiment, along with the fractional change in $C_L^{\Phi \Phi}$ for
some fiducial values of $\sum m_\nu$ relative to the $\sum m_\nu = 0$
case.

\subsection{Tomographic galaxy clustering with spectroscopic surveys}

\label{galaxyforecast}

Starting in 2014, the Extended Baryon Oscillation Spectroscopic Survey
(eBOSS) will use the BOSS spectrograph to perform spectroscopy on a
massive sample of galaxies and quasars in the redshift range that lies
between the BOSS galaxy sample and the BOSS Lyman-$\alpha$ sample.
The targets for eBOSS spectroscopy will consist of Luminous Red
Galaxies (LRGs: $0.6<z<0.8$), Emission Line Galaxies (ELGs:
$0.6<z<1.0$), ``clustering'' quasars to directly trace large-scale
structure ($1<z<2.2$), and Lyman-$\alpha$ quasars ($2.2<z<3.5$).  The
comoving volume probed by eBOSS will be nearly ten times that probed
by the BOSS galaxy survey (though at lower sampling density). 

The success of BOSS not only led to the eBOSS program, but has largely
inspired the concept of DESI\@.  DESI is a proposed wide-field
spectroscopic survey, to be conducted on a 4-m class telescope to
study dark energy.  DESI represents one of the top priorities for new
Office of Science Cosmic Frontiers efforts as identified in the recent
``Rocky III'' report; CD-0 approval was granted in September 2012.
The preliminary DESI survey design covers 14,000 deg$^2$ with
spectroscopic observations of LRGs ($0.6<z<1.0$), ELGs ($0.6<z<1.5$),
and quasars ($1<z<3.5$) for direct clustering and Lyman-$\alpha$
forest.  The density and effective volume probed by DESI will far
exceed either BOSS or eBOSS. The details of the survey specifications
are provided in Ref.~\cite{Font-Ribera:2013rwa}.

As a natural consequence of large spectroscopic BAO programs such as
BOSS, eBOSS, and DESI, clustering in the density field is sampled
across a wide range of $k$ modes over a very large volume.  As
explained in Section~\ref{physicsofnucosmo}, the signature of
neutrinos appears as a characteristic suppression of power below
certain scales in the matter power spectrum.  Effort within the BOSS
collaboration has primarily centered on reconstructing the broad-band
power spectrum to extract information about BAO, redshift space
distortions, and bias parameters from galaxies and Lyman-$\alpha$
forest.  Continued improvements on estimates of the power spectrum
will soon allow tests of neutrino masses with the BOSS data.

Assuming a baseline of Planck CMB measurements, we forecast for
sensitivity for neutrino mass with BOSS, eBOSS and DESI
\cite{Font-Ribera:2013rwa}.  We assume measurements of large-scale
modes with wavelengths up to $k_{\rm max}=0.1\ h\rm\ Mpc^{-1}$
(conservative estimates) and $k_{\rm max}=0.2\ h\rm\ Mpc^{-1}$
(optimistic estimates).\footnote{These constraints are somewhat more
  conservative that those used in \cite{Ade:2013zuv}, where $k_{\rm
    max}$ of $0.15\ h\rm\ Mpc^{-1}$ and $0.3\ h\rm\ Mpc^{-1}$ were
  used as conservative and optimistic estimates respectively.}
Results can be found in the Table~\ref{tab:tab}.

For the eBOSS experiment, we predict a 68\% measurement error of
$\sigma(\sum m_{\nu}) = 36/52\rm\ meV$ for optimistic/conservative
estimates when combining the results from the eBOSS LRGs, ELGs and
quasars.  The upper limit on neutrino masses derived from eBOSS is
comparable to the minimum allowed mass in an inverted hierarchy.
Current constraints on the effective number of neutrino species of
$N_{\rm eff} =3.30 \pm 0.27$ from BAO and Planck CMB data will be
improved by a factor of two by including eBOSS measurements.  DESI
should improve these constraints considerably. In particular, neutrino
masses can be constrained to $\sigma(\sum m_\nu)=17/24\rm\ meV$
(optimistic/conservative) from galaxies alone, enough to make a
measurement of the minimum neutrino mass at $\sim 3\sigma$ confidence.

\subsection{Tomographic Weak Lensing in Galaxy Surveys}

\label{weaklensingforecast}

Recent determinations of the sensitivity of upcoming ground based
surveys such as the Large Synoptic Survey Telescope
(LSST)~\cite{Ivezic:2008fe} have a forecast sensitivity of deep
background galaxy imaging at the level of $\sigma(\sum m_\nu) = 23\rm\
meV$~\cite{Joudaki:2011nw}.  This value includes the tomographic
galaxy clustering information that would also be achieved from LSST
covering one-half of the sky, an average galaxy density of 50
arcmin$^{-2}$, using methods discussed in \S\ref{galaxyforecast}.  A
space-based mission with wide field of view and covering 10\% of the
sky with 100 arcmin$^{-2}$ sensitivity and mean lensed galaxy redshift
of $\langle z\rangle = 1$ was found to have the sensitivity of
$\sigma(\sum m_\nu) = 23\rm\ meV$ as well.  Such a survey may be
achieved by some configurations considered for the proposed Wide-Field
Infrared Survey Telescope-Astrophysics Focused Telescope Asset
(WFIRST-AFTA)~\cite{Spergel:2013tha}.  Both of these forecasts
employed simultaneous Planck-level determinations of cosmological
parameters, but allowed for extra parameter variation, including
$\nnu$ and running of the primordial power spectrum
index~\cite{Joudaki:2011nw}.  Relaxing from a non-minimal cosmological
model, as discussed in \S\ref{theorypriors} relaxes these constraints.

The upcoming European Space Agency {\sc Euclid} mission will also
perform a deep galaxy survey and weak lensing
survey~\cite{Laureijs:2011gra}.  Planck CMB plus {\sc Euclid} weak
lensing has been forecast to have a sensitivity of $\sigma(\sum m_\nu)
= 25\ \rm meV$ in a minimal cosmological model, with the upper limit
depending on assumptions on galaxy
bias \cite{Carbone:2010ik,Hamann:2012fe}.  Forecasts of a combination
of angular and tomographic weak lensing shear auto-correlation, galaxy
auto-correlation, and galaxy-shear cross-correlation sample data from
{\sc Euclid} has been forecast to have a sensitivity of $\sigma(\sum
m_\nu) = 23\rm\ meV$, when including a marginalization over the galaxy
bias model \cite{Hamann:2012fe}.  The photometric galaxy cluster
catalog would also have sensitivity to neutrino
mass~\cite{Carbone:2011by}. It has been found in
Ref.~\cite{Basse:2013zua} that $\sigma(\sum m_\nu)$ could improve to
$10$ meV range with the full combination of weak lensing, galaxy
clustering, galaxy-shear cross-correlation, in combination with a new
a weak-lensing selected galaxy cluster sample derived from the {\sc
Euclid} galaxy survey.  However, this sensitivity would require large
improvements in our understanding of observational systematics of all
of these combinations of probes.

\subsection{Other Probes: Galaxy cluster surveys, Lyman-$\alpha$ Forest, and 21 cm Surveys}
\label{otherlssforecast}

There have been forecasts for a number of other tests of cosmological
LSS as a probe of neutrino masses.  These include the clustering of
neutral gas as probed in the Lyman-$\alpha$ forest, galaxy cluster
samples, and 21 cm surveys.  These methods also have complicated
relations between the observables and inferred linear primordial LSS
clustering, yet the inferred sensitivity on neutrino mass from these
observations may be competitive with other probes.

The Lyman-$\alpha$ forest is the pattern of absorption features in
quasar spectra due to intervening low density gas.  This gas is
expected to follow gravitationally-dominant dark matter clustering in
the linear to weakly non-linear regime of clustering.  Going from the
observed flux distribution to the power spectrum of matter in LSS
requires knowing the {\it bias} of gas to matter, which includes
knowing the temperature-density relation of the gas and its evolution
over cosmic history, as well as the nature of the ionizing background
radiation impinging on the gas.  Setting aside the systematic
uncertainties, the Lyman-$\alpha$ forest has some of the strongest
inferred constraints on neutrino mass: using results from WMAP 3-year
CMB observations with SDSS quasar's Lyman-$\alpha$ forest, the upper
limit on the neutrino mass is $\sum m_\nu < 0.17\rm\ eV$ (95\%
C.L.) \cite{Seljak:2006bg}. We note however, that the calibration of
the absolute power amplitude in the subsequent WMAP releases has
changed in the direction that would relax these limits. The Planck
absolute power measurements are even higher that the final WMAP
measurements, further weakening these constraints.  Follow-up analyses
found that it is possible to constrain neutrinos even in the absence
of CMB data prior giving an upper limit of $\sum m_\nu < 0.9\rm\ eV$
(95\%) based on SDSS quasar's Lyman-$\alpha$
forest \cite{Viel:2010bn}.  Optimistic forecasts for high-resolution
Keck and/or VLT spectra of the Lyman-$\alpha$ forest combined with
Planck-level CMB constraints find their reach to be $\sum m_\nu <
0.11\rm\ eV$ (95\% C.L.)~\cite{Gratton:2007tb}.  Modeling of the
Lyman-$\alpha$ forest through numerical simulations is a very active
field with several U.S. and European groups currently doing
research. It is likely that the future constraints will be limited by
the systematic accuracy with which we can make these theoretical
predictions.

An upcoming probe of the high-redshift gas is low-frequency radio
observations of the redshifted 21~cm line of neutral hydrogen, which
has the potential to map the matter distribution through the epoch of
reionization, out to $z\sim 12$.  A key requirement of 21~cm surveys
is the removal of Galactic synchrotron emission, which is four to five
orders of magnitude greater than the 21~cm signal from high-$z$.  The
smoothness of the foreground has been shown to be usable to overcome
its dominance and 21~cm clustering measurements out to $z\sim 0.8$
have been achieved~\cite{Switzer:2013ewa}.  Because of the potential
of mapping neutral hydrogen to high-redshifts, large volumes of matter
clustering may be measurable, revealing an unprecedented number of
$k$~modes~\cite{Morales:2009gs}.  Several 21~cm experiments are
operational or being developed, including the LOw Frequency ARray
(LOFAR)\footnote{\tt http://www.lofar.org/},the Precision Array for
Probing the Epoch of Reionization (PAPER)\footnote{\tt
http://astro.berkeley.edu/$\mathtt \sim$dbacker/eor/}, the Murchison
Widefield Array (MWA)\footnote{\tt http://www.mwatelescope.org/}, and
the Canadian Hydrogen Intensity Mapping Experiment
(CHIME)\footnote{\tt http://www.physics.ubc.ca/chime/}.  The highest
sensitivities would come from a large-scale 21~cm survey facility
proposed as the Square Kilometer Array (SKA)\footnote{\tt
http://www.skatelescope.org}.  For example, forecasts for MWA
sensitivities to neutrino mass, with Planck CMB constraints, are
$\sigma(\sum m_\nu) = 27\rm\ meV$, and for SKA plus Planck reach
$\sigma(\sum m_\nu) = 17\rm\ meV$ in optimistic scenarios of
ionization modeling, but can decrease in sensitivity by a factor of
$\sim$3 or more for less optimistic modeling
scenarios \cite{Mao:2008ug}.

Galaxy clusters are the most massive bound structures in the universe,
and their abundance is highly sensitive to the amplitude and shape of
the matter power spectrum on scales of $k \approx 0.1\ {\rm Mpc}^{-1}
h$.  Galaxy cluster samples have been created from X-ray surveys
detecting intra-cluster gas emission, optical surveys of member galaxy
clusters, and millimeter-wave surveys which detect the
Sunyaev-Zeldovich (SZ) effect from inverse Compton scattering.  A
limiting systematic in galaxy-cluster based measures of the primordial
power spectrum and therefore neutrino mass is the connection between
the observable quantity and the inferred galaxy cluster halo mass, the
latter of which is predicted by structure formation simulations.
Forecasts for weak-lensing-selected galaxy clusters from an LSST-like
surveys are optimistic with sensitivities of $\sigma(\sum m_\nu) =
30\rm\ meV$~\cite{Wang:2005vr}.  Optically-selected galaxy cluster
samples can improve on the constraints from galaxy-clustering and
weak-lensing measures, as discussed in \S\ref{weaklensingforecast}.

\section{Systematics}

\subsection{Theoretical Priors}
\label{theorypriors}

The goal of constraining and potentially detecting neutrino mass and
neutrino number density cosmologically is based within a rigorously
constrained and well-tested model.  However, that model has some
inherent theoretical priors and simplifying assumptions.  Almost all
of the constraints on neutrino mass discussed here and in the
literature are constraints on neutrinos as an extension to the
standard cosmological model, minimally described by six parameters, but
sometimes extended to up to 10 parameters.  A detailed
discussion of the model dependence of the cosmological constraints on
neutrino mass was given in Ref.~\cite{Abazajian:2011dt}, and we
summarize some of the model dependencies here.  As discussed in the
introduction, if a non-minimal model is detected robustly by
cosmological observations, with $\nnu \neq 3.046$ and $\sum m_\nu \gg
58\rm\ meV$, this opens the possibility of an indication for other
extensions to cosmology, including curvature, non-constant dark
energy, a non-uniformly scale invariant primordial perturbation
spectrum, extra particle or radiation species, and even other
possibilities.  Conversely, if the minimal model is detected with high
significance, that detection rests on the theoretical edifice of
modern cosmology, albeit a successful model.  If laboratory
experiments of neutrino mass and properties are at odds with the
cosmological results, then the implications for cosmological theory
would necessarily be addressed.

Other theoretical assumptions are also present even within the
10-parameter models that are not commonly explicitly stated.  These
include the assumption that the primordial power spectrum arising out
of inflation is a power-law, $P(k) \propto k^n$, or, in extended
models at most having a curvature ``rolling'' of the power law,
$P(k) \propto k^{n + (dn/d\ln k) (k/k_0)}$.  There has been some work
that shows that dropping the assumption of a nearly scale-independent
power law spectrum could provide cosmological observations that have
drastically different neutrino sectors, but produce the same CMB
observations~\cite{Kinney:2001js}.  Other work has shown that
constraints on cosmological parameters expand considerably when
including arbitrary yet continuous functions for the initial
conditions of the primordial perturbation
spectrum \cite{Hazra:2013eva}.  That work did not address neutrino
constraints specifically, but presumably neutrino mass constraints
would also greatly degrade once the assumption of the primordial
spectrum is relaxed similarly.  It should be emphasized that
deviations from a primordial nearly scale-invariant spectrum are not
predicted by standard inflation models~\cite{Bassett:2005xm}.  Other
implicit assumptions include that Einstein's gravity is valid on
galaxy to cosmological scales (e.g.,~\cite{Ishak:2005zs}).

Note that there also exist model-dependent assumptions in some of the
LSS observations' interpretations leading to the constraints described
in the previous sections.  These include the nature of galaxy biasing
with respect to dark matter clustering in the galaxy auto and
cross-correlation statistics studied, e.g, in
Ref.~\cite{Basse:2013zua}.  In either optically selected or weak
lensing selected cluster-counting constraints, the mass function of
halos and the mass-observable relation have inherent model
uncertainties that are only partially quantified, at this point.  In
addition, some of the high angular resolution weak lensing shear
constraints use information from where weak lensing has the highest
signal-to-noise, at weak-lensing $\ell^{\rm WL}$ multipoles where
matter clustering is in the deeply nonlinear regime.  This would
require an accurate and precise prediction of the nonlinear matter
power spectrum~\cite{Heitmann:2013bra} including significant effects
on matter clustering from
baryons~\cite{Zhan:2004wq,Jing:2005gm,Zentner:2007bn} and even
neutrino clustering~\cite{Abazajian:2004zh,Brandbyge:2008rv,VillaescusaNavarro:2012ag,LoVerde:2013lta,Rossi:2014wsa}.

\subsection{CMB Foregrounds and Systematics}

Reconstruction of the matter deflection field from observations of the
lensed CMB -- and, by extension, CMB lensing constraints on neutrino
physics -- rely on the non-Gaussian nature of the lensed CMB; the
deflection of CMB photon trajectories by the intervening gravitational
potential induces correlations between initially independent spatial
modes, and this permits the potential to be reconstructed. Because
astrophysical foregrounds are non-Gaussian they can affect lensing
reconstruction in a complicated way.  Similarly complicated will be
the way in which instrumental polarization systematics come into the
higher-order calculations involved in lensing reconstruction. Fully
quantifying these effects will be an involved task that will almost
certainly require end-to-end simulations.

For the purpose of this Community Study, we assess the effects of
foregrounds and systematics on lensing reconstruction in a simplified
way.  Since $B$-mode polarization is by far the faintest signal used
in lensing estimators, the foregrounds and systematic errors in
lensing reconstruction have the largest impact on measurements of
$B$-mode polarization.  If we simply ensure that the false $B$ modes
from astronomical and systematic residuals are smaller than the noise
level, their effects on the much larger $E$-mode signal will be
completely negligible, and lensing reconstruction will be
noise-limited.  This significantly simplifies the considerations of
frequency coverage and instrument specifications, because predicting
the levels of astronomical and instrumental $B$-mode polarization is
much more straightforward than dealing with higher-order statistics.
In addition, a large volume of literature already exists on the
forecast and mitigation techniques, thanks to the global pursuit of
tensor-mode perturbations
(e.g.,~\cite{Hu:2002vu,MacTavish:2007kh,Takahashi:2009vp}). $B$-modes
from primordial gravitational waves are at larger scales, smaller
$\ell$'s, and do not largely affect the lensing $B$-mode
reconstruction of the scales of interest~\cite{Kaplinghat:2003bh}.

\subsubsection{CMB Foregrounds: Polarized Point Sources}
Focusing on the level of residuals in $B$-mode measurements, we can
quickly conclude that polarized extragalactic point sources are
expected to be the largest contaminant of the reconstructed potential
\cite{Smith09}.  The $B$-mode power spectrum from point sources is
dominated by bright sources, which can be identified and masked up to
a flux limit.  The level of residual $B$ modes after source cuts
depends on the degree of polarization of the sources.  The mm-wave
flux density distributions of both radio sources and dusty galaxies
have been well established in recent years
\cite{Vieira:2009ru,Marriage:2010yy,Ade:2011bb}. The polarization
properties of these sources at mm wavelengths is less well-known, but
strong upper limits can be placed on the mean-squared polarization
fraction of both source families \cite{Battye:2010zz,
  Seiffert:2006vh}.  Using these upper limits as maximally pessimistic
assumptions for foreground contamination, robust predictions can be
made for CMB-S4.  In the 2500 $\mathrm{deg}^2$ SPT-SZ survey, with a
$\sim 6$~mJy cut, the residual source rms is already below
$10\ \mu$K-arcmin in temperature.  Since the expected source
polarization is 2-5\%, the residual rms in polarization will be
significantly below $1 \mu$K-arcmin, the benchmark sensitivity for
CMB-S4.  At 95 GHz and 150 GHz, there is approximately one source per
square degree with a flux density greater than $6$~mJy.  Therefore,
they can be easily identified and removed with minimal data loss using
internal CMB-S4 data (a $6$~mJy source will be detected at $\sim 30
\sigma$ for a 95~GHz experiment with $\sim 1 \mu$K-arcmin and $3'$
FWHM).

\subsubsection{CMB Instrumental Systematics}
Tight control of all potential sources of systematic contamination is
required for any deep measurement of faint $B$-mode polarization
signals.  Since lensing reconstruction is ultimately limited by the
amount of false $B$-mode signal, the methodology developed for
predicting and mitigating instrumental systematics for inflationary
$B$ modes can be directly used to estimate the experimental
specifications for CMB lensing.  Beam imperfections, gain
uncertainties, and polarized sidelobe pickup can all convert much
brighter temperature fluctuations to false $B$ modes.  Errors in
polarization angle calibration or telescope pointing can falsely mix
brighter $E$ modes into $B$.

Drawing on the extensive literature on polarization systematics, we
conclude that the most potentially dangerous instrumental sources of
spurious $B$-mode polarization are gain and beam errors that mix $T$
into $B$.  For example, a gain mismatch between detectors measuring
two orthogonal polarizations directly converts $T$ anisotropy to
polarization (``monopole leakage'').  The required level of gain
control for lensing science can be estimated by simply requiring the
$T$ leakage be smaller than the noise level in $B$.  For CMB-S4 noise
levels ($\sim 1\mu$K-arcmin), we estimate that if gain mismatch is
controlled at the $< 10^{-3}$ level, the false $B$ signal will be less
than the noise level for $\ell>300$, where most of the lensing
information resides.  This level of relative gain calibration is
challenging, but the requirement can be significantly alleviated by
projecting measured $T$ modes out of polarization maps. This technique
removes any bias from $T \rightarrow B$ monopole leakage, at the cost
of a slight increase in variance. The required level of relative gain
calibration is relaxed to $10^{-3}\xi$, where $\xi$ is the
signal-to-noise ratio of the $T$ template.  Since $\xi$ is easily in
the hundreds with CMB-S4 $T$ maps, the required calibration precision
will become very manageable.

A beam mismatch between detectors measuring orthogonal polarizations
also converts $T$ anisotropy to a false polarization signal, but with
some dependence on angular scale (as opposed to the monopole leakage
from gain mismatch). Because the dominant modes of this systematic are
a dipole and quadrupole at the scale of the beam, the effect strongly
mitigated when the instrument beam size is smaller than the smallest
scientific scale of interest.  For example, a $3'$ FWHM beam places
the peak of the quadrupole beam-mismatch effect at $\ell \sim 15,000$,
an order of magnitude above the peak of the lensing $B$-mode spectrum.

\subsection{Deviations from Uniformity of Density Tracers in LSS}
\label{lss-systematics}
The Fisher matrix projections for neutrino mass constraints from
spectroscopic LSS surveys assume that the large-scale clustering
measurements will be limited by statistical errors.  This requires
stringent control of systematic errors such as incomplete modeling of
astrophysical effects and systematic errors arising from instrument
performance and survey design.

The astrophysical systematics arising from scale-dependent biasing are
perhaps not such a big problem as one would naively expect, since
neutrino suppression creates a very specific distortion of the matter
power spectrum.  At the scales of interest, other effects will receive
perturbative corrections in even powers of $(k/k_x)$, where $k_x$ is
the relevant small scale (non-linear scale for non-linear clustering,
locality scale for biasing, etc.), using methods for non-linear
modeling as discussed, e.g., in
Refs.~\cite{Cole:2005sx,Eisenstein:2006nj}

On the other hand, experience with BOSS has found that observational
systematics can be significant.  Contamination arises because
intrinsic fluctuations on the largest scales are small and systematics
effects can modulate the data on arbitrarily large scales.  Examples
of such contamination include the impact of stellar contamination and
dust extinction on target selection efficiency, variations in seeing
during imaging that alter target selection, varying atmospheric
conditions in spectroscopy that effect redshift success, and so on.

BAO measurements are largely protected from such effects on data
quality because they rely on a relatively sharp feature, but
constraints on neutrino masses require a well calibrated measure of
absolute power as a function of wavenumber.  These effects are
therefore important primarily for neutrino mass estimates and
constraints on the effective number of neutrino species because they
use the full shape of the galaxy power spectrum.

These systematics have already been extensively studied within BOSS
\cite{Ross:2011cz, Ross:2012qm, Ho:2012vy,Pullen:2012rd}, and the
greater volume and greater statistical power at large scales from
eBOSS will place new demands on homogeneity of the target samples.  At
this point in time, the eBOSS target selection is being developed and
tested for systematic effects arising from non-uniform imaging data.
The primary tests conducted to date include measures of the angular
correlation function and assessment of uniformity as a function of
imaging data quality.  In addition, more advanced surveys such as
eBOSS will utilize mixed tracers of the density field in overlapping
regions, allowing cross-correlation analysis as a probe of systematic
trends in any single target class.  Just as the lessons from BOSS have
informed eBOSS target selection, eBOSS observations will be an
essential step in preparation of target selection and survey design
for DESI.

\section{Conclusions}

The experimental quantification of the cosmological neutrino
background has achieved the robust detection of the neutrino
background energy density through its effects on the CMB and LSS, and
the measures of properties of neutrinos in the cosmological background
have achieved unprecedented precision and accuracy. These measures
compete in precision with laboratory probes of neutrino number and
mass, albeit with the theoretical priors discussed in
\S\ref{theorypriors}.  The current generation of Planck CMB with
high-$\ell$ CMB, SDSS, BOSS and 6dF LSS data have constrained the sum
of neutrino masses to be $\sum m_\nu < 0.23\rm\ eV$ (95\% C.L.) and
number of neutrinos to be $N_{\rm eff} =3.30 \pm 0.27$.

We have reviewed the planned and proposed future experiments that
would be more sensitive to cosmological neutrinos, including galaxy
surveys, weak lensing surveys, other LSS observations, as well as a
Stage-IV CMB experiment.  We find that the experiments individually
achieve optimal sensitivities to the sum of neutrino masses at the
level of $\sim 20-30\rm\ meV$, while combined CMB plus LSS probes
achieve greater sensitivities.

Much of the sensitivity to $\sum m_\nu$ comes from measurements of the
gravitational lensing of the CMB and from measurements of galaxy
clustering. Significant progress has been made recently in both of
these areas, with the first projected mass reconstructions from CMB
lensing
\cite{Das:2011ak,vanEngelen:2012va,Das:2013zf,Holder:2013hqu,Ade:2013tyw,Geach:2013zwa},
detection of the CMB lensing $B$-mode polarization
\cite{Hanson:2013daa}, and percent level measurements of the distance
scale from BAO measurements \cite{Blake:2012pj,Anderson:2012sa,
  Busca:2012bu,Slosar:2013fi}.

We find that future combined probes of a Stage-IV CMB experiment with
BAO information from the MS-DESI galaxy survey can achieve
\begin{eqnarray}
&&\sigma\left(\sum m_\nu\right) = 16\rm\ meV\, ,  \nonumber\cr
&&\sigma\left(\nnu\right) = 0.020\, ,  \nonumber
\end{eqnarray}
in the case of a standard six parameter $\Lambda$CDM model extended by
the respective neutrino parameter.

Discrimination between changes to neutrino properties and other
changes to the standard cosmological model will be supported by the
diversity of very specific observable differences generated by changes
to neutrino properties.  For example, increasing $N_{\rm eff}$ not
only suppresses small-scale fluctuation power, but also leads to
detectable change to the temporal phase of the acoustic oscillations.
Increasing $\sum m_\nu$ in a manner that is consistent with the CMB
power spectrum leads to very specific redshift-dependent changes to
the expansion rate that change sign at $z \sim 1$, as well as specific
scale and redshift-dependent changes to the power spectrum of the
galaxy distribution.  In short, neutrinos leave a very specific
imprint on various observables and these signatures will provide
an important battery of consistency checks.

Given that the minimal normal-hierarchy model of neutrino mass
requires a sum of neutrinos of at least 58 meV, and a standard thermal
history requires $\nnu = 3.046$, then experimental and observational
neutrino cosmology can move beyond a detection of the neutrino
background, where we are today, to a precise detection of the
``fingerprints'' of both neutrino mass and number in the standard
cosmological model. Complementarily, a robust detection away from the
standard predictions would be a definitive detection of new physics.

\newpage

\section*{Appendix:\, \, A Stage-IV CMB experiment, \cmbexp}
\label{Appendix:CMB-S4}

\cmbexp\ is a Stage-IV CMB experiment with two goals: to search for
inflationary B-modes and to measure the sum of neutrino masses to an
accuracy of 10-15 meV at 68\% confidence. It occupies a unique
position in the HEP science portfolio as the measurement of CMB
polarization is the only method known to probe high-scale physics.  As
discussed in this document, measurement of CMB polarization provides a
powerful and complementary approach to understanding the neutrino
masses. The Stage IV \cmbexp\ experiment is the logical next step for
the U.S.\ world-leading ground-based CMB program. It will provide the
order of magnitude increase in sensitivity required to reach the goal
of $\sigma(\Sigma m_\nu) = 16 {\rm meV}$ (when combined with data from
DESI) and an uncertainty in the tensor to scalar ratio of $\sigma(r) =
0.001$.

To achieve these goals, \cmbexp\ will observe 50\% of the sky to the
unprecedented noise level of $\le 1 \mu \mathrm{K}$ per arcminute
pixel with an angular resolution of $\le 3$ arcminutes. At higher
angular resolutions the sensitivity of \cmbexp\ to secondary CMB
anisotropy would greatly expand its science reach, e.g., through
probing Dark Energy and $\Sigma m_\nu$ through Sunyaev-Zel'dovich (SZ)
cluster cosmology.  Reaching noise levels of $\le 1 \mu
\mathrm{K}$-arcminute is an ambitious undertaking requiring of order
500,000 background limited detectors on the sky with broad spectral
coverage over 40 - 240 GHz, and integrating for several years.  The
detectors would be distributed across multiple platforms at two or
more sites to provide large sky coverage and optimized optical
throughput. Thus, realizing \cmbexp\ requires reshaping the HEP CMB
program from its current state of fragmented small efforts into a
coherent well-supported experimental program.

\subsection*{Detector Technology for \cmbexp}

\begin{figure}[t]
\centering \includegraphics[width=0.8\textwidth]{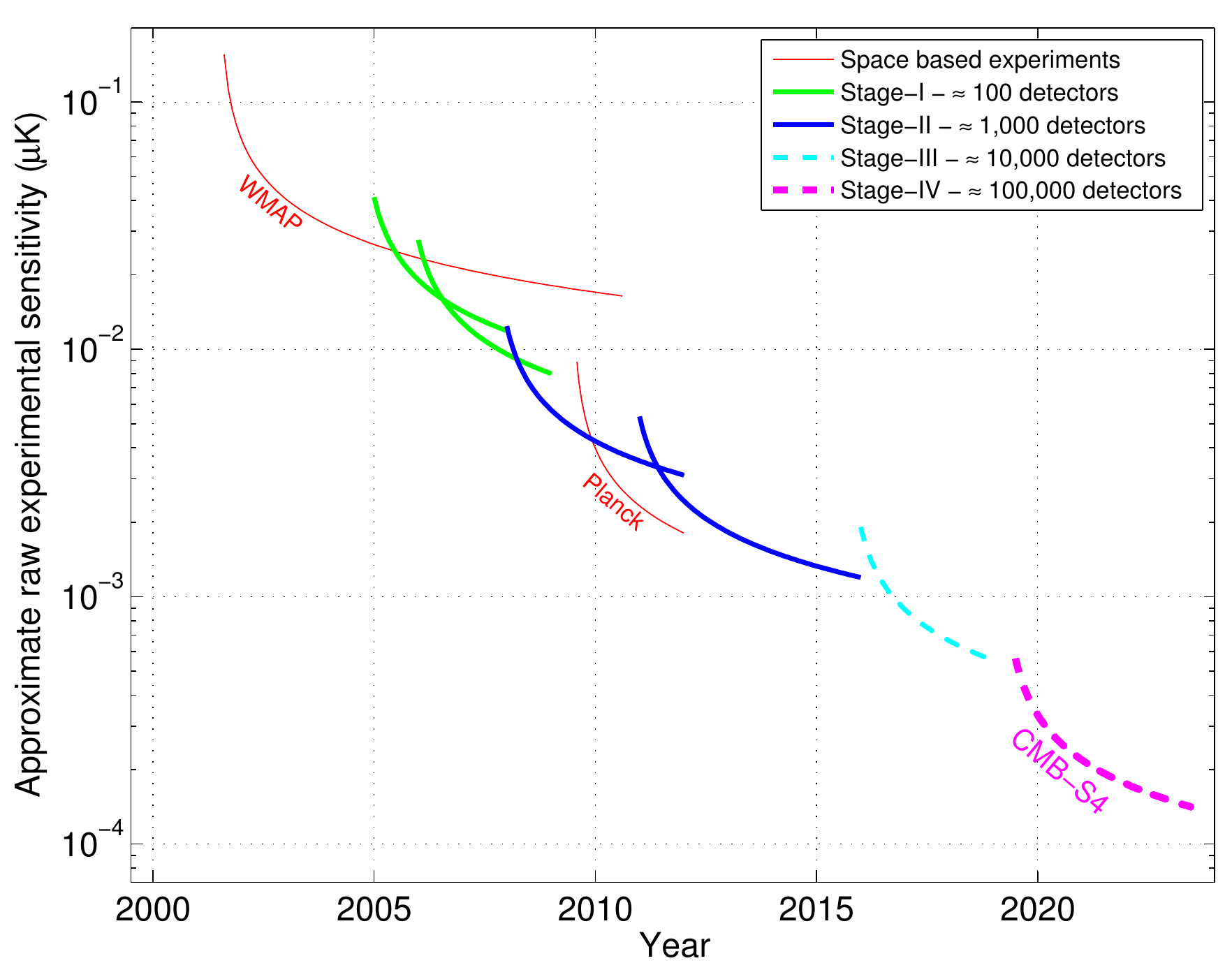}
\caption{Plot illustrating the evolution of the raw sensitivity of CMB
  experiments, which scales as the total number of
  bolometers. Ground-based CMB experiments are classified into Stages
  with Stage II experiments having $O$(1000) detectors, Stage III
  experiments having $O$(10,000) detectors, and a Stage IV experiment
  (such as \cmbexp) having $O$(100,000) detectors.}
\label{fig:expt_progress}
\end{figure}

The technical requirements for \cmbexp\ are dictated by the
fundamental limitations of CMB measurements.  Specifically, all
competitive CMB detectors are sensitivity-limited where the dominant
noise in an individual detector element comes from shot noise arising
from the arrival time of the photons. Thus, achieving the required
\cmbexp\ sensitivity requires increasing the number of detected modes,
which is straightforward to achieve by increasing the number of
detectors (see Fig.~\ref{fig:expt_progress}). \cmbexp\ will have 500
times more detectors than the current state-of-the-art Stage II
experiments, 30 times more than planned Stage III experiments, making
scaling the primary technical challenge of \cmbexp.

\begin{figure}[t]
\centering
\includegraphics[trim=0.7in 0 1.5in 0,clip,scale=0.3]{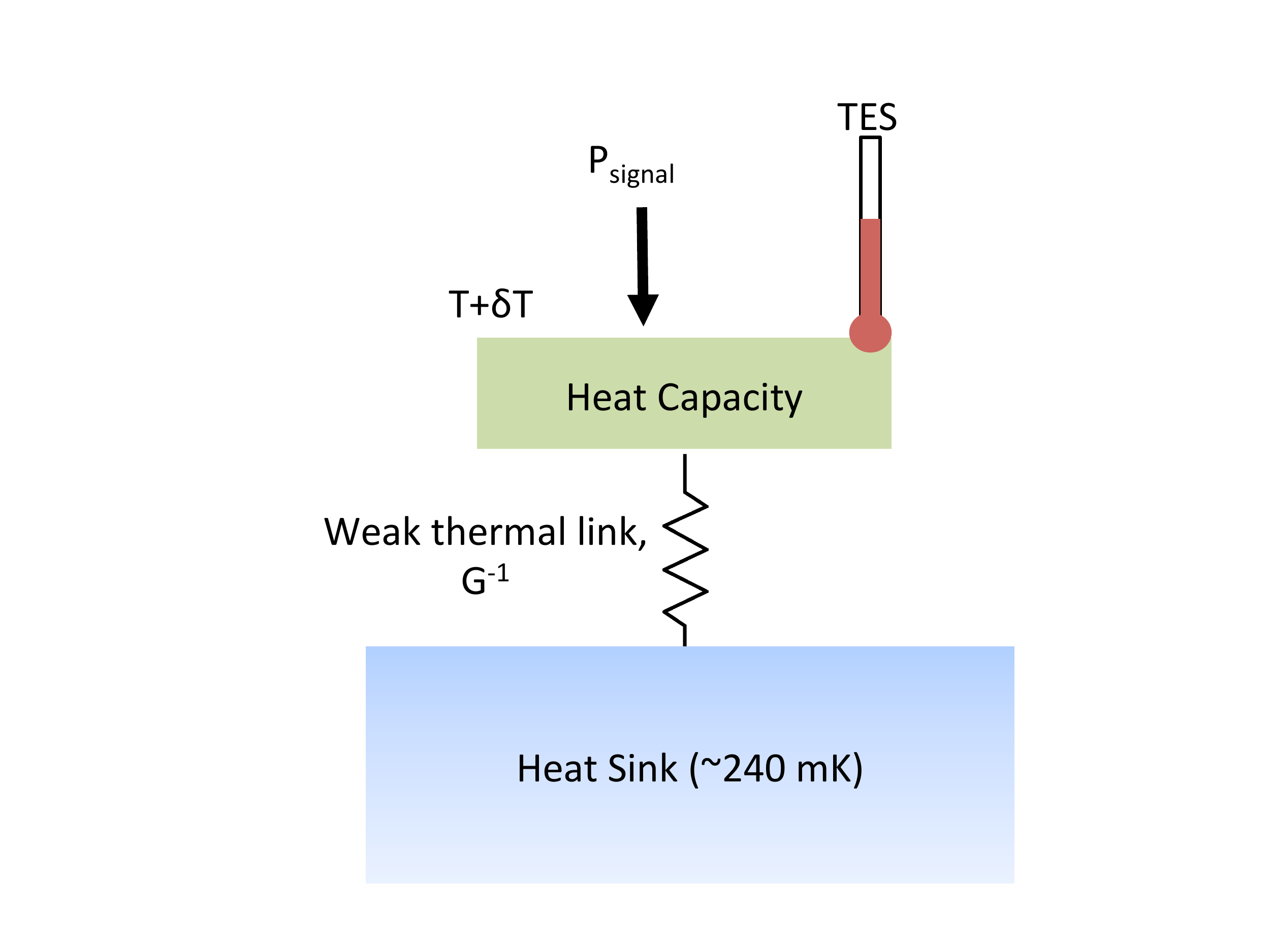}
\includegraphics[trim=0.4in 0.4in 0.1 0,clip,scale=0.23]{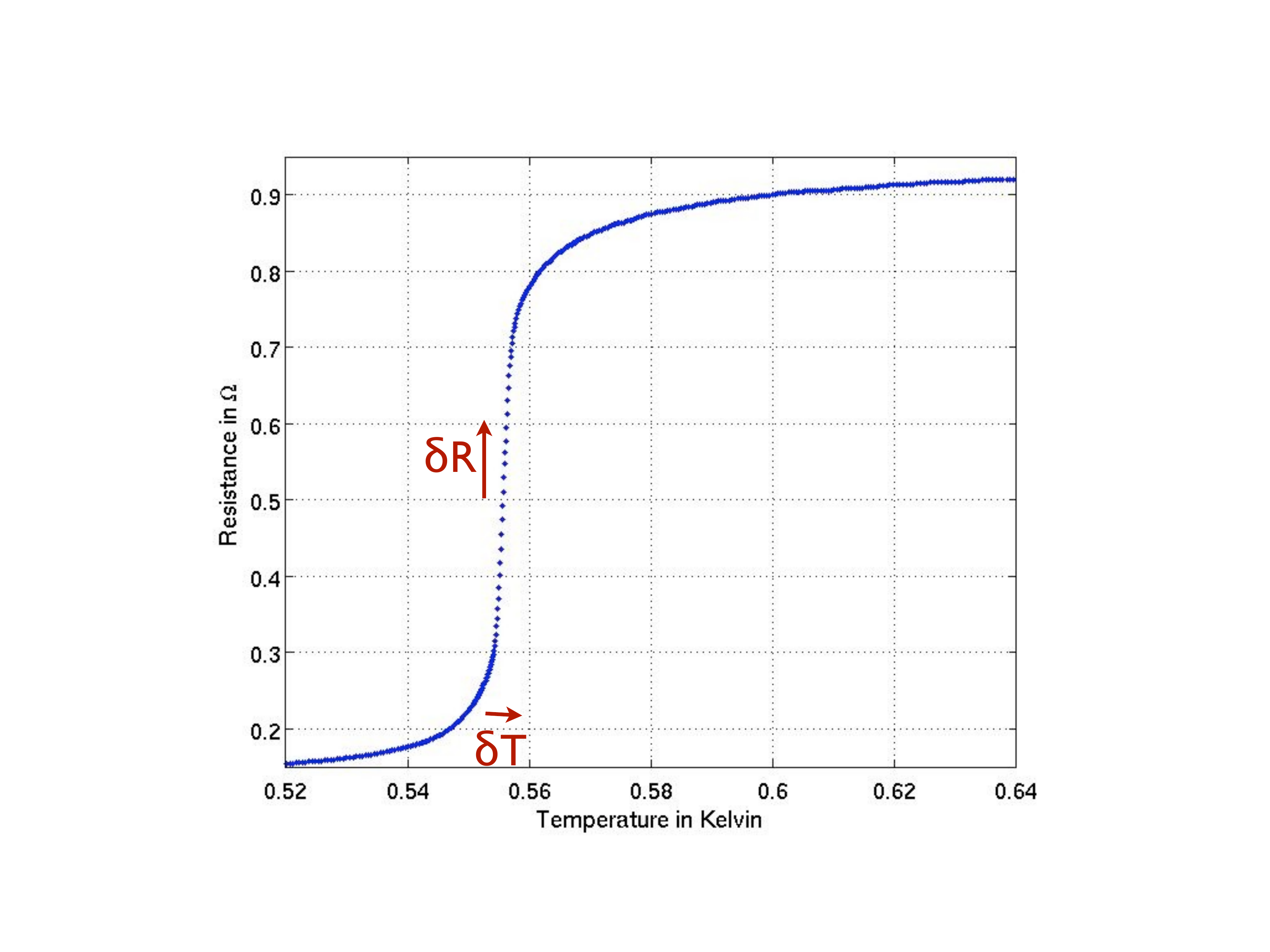}
\vskip 0pt
\caption{Left: Illustration of a thermal circuit for a typical
  Transition Edge Sensor (TES) detector highlighting the principles of
  signal detection. A weakly thermally sunk heat capacity absorbs
  power, P$_{\rm signal}$, which is to be measured. Variations in the
  absorbed power change the heat capacity's temperature, which is
  measured by a TES operating under strong electro-thermal
  feedback. Right: Plot of resistance versus temperature for a typical
  TES illustrating the principles of negative electro-thermal
  feedback~\cite{irwin:1998}. The TES is voltage biased into the
  middle of its superconducting-to-normal transition. Small changes in
  the TES temperature produce large changes in the TES
  resistance. Since the TES is voltage biased, an increase (or
  decrease) in the temperature produces an increase (or decrease) in
  the resistance leading to a decrease (or increase) in the Joule
  heating power supplied by the bias. This canceling effect
  corresponds to a strong negative electro-thermal feedback making the
  current through the TES nearly proportional to P$_{\rm signal}$.}
\label{fig:TEScartoon}
\vskip -12pt
\end{figure}

Towards this end, \cmbexp\ will utilize Transition Edge Sensor (TES)
bolometers as its baseline detector technology.\footnote{We note that
  at the lowest frequencies envisioned, $\sim40$ GHz, MMIC-amplifier
  or new superconducting amplifier technologies may be practical.} A
TES is an ultra-sensitive thermometer consisting of a thin
superconducting film weakly heat-sunk to a bath temperature much lower
than the superconductor $T_{c}$ (see Fig.~\ref{fig:TEScartoon},
left). The principles of operation are simple to understand. By
supplying electrical power to the TES, we can raise the temperature of
the sensor so that the film is in the middle of its
superconducting-to-normal transition (see Fig.~\ref{fig:TEScartoon},
right). If the electrical power is supplied via a voltage bias, a
negative feedback loop is established~\cite{irwin:1998}. Small changes
to the TES temperature, arising from thermal fluctuations (noise) or
changes in the absorbed power from a source (signal), lead to large
changes in the TES resistance. The change in resistance creates a
canceling effect because increases (or decreases) in temperature
produce decreases (or increases) in Joule heating power. This negative
electro-thermal feedback is very strong because the transition is very
sharp. It linearizes the detector response and expands the detector
bandwidth.

\begin{figure}[t!]
\centering
\includegraphics[trim=0 0.75in 0 .75in,clip,scale=0.3]{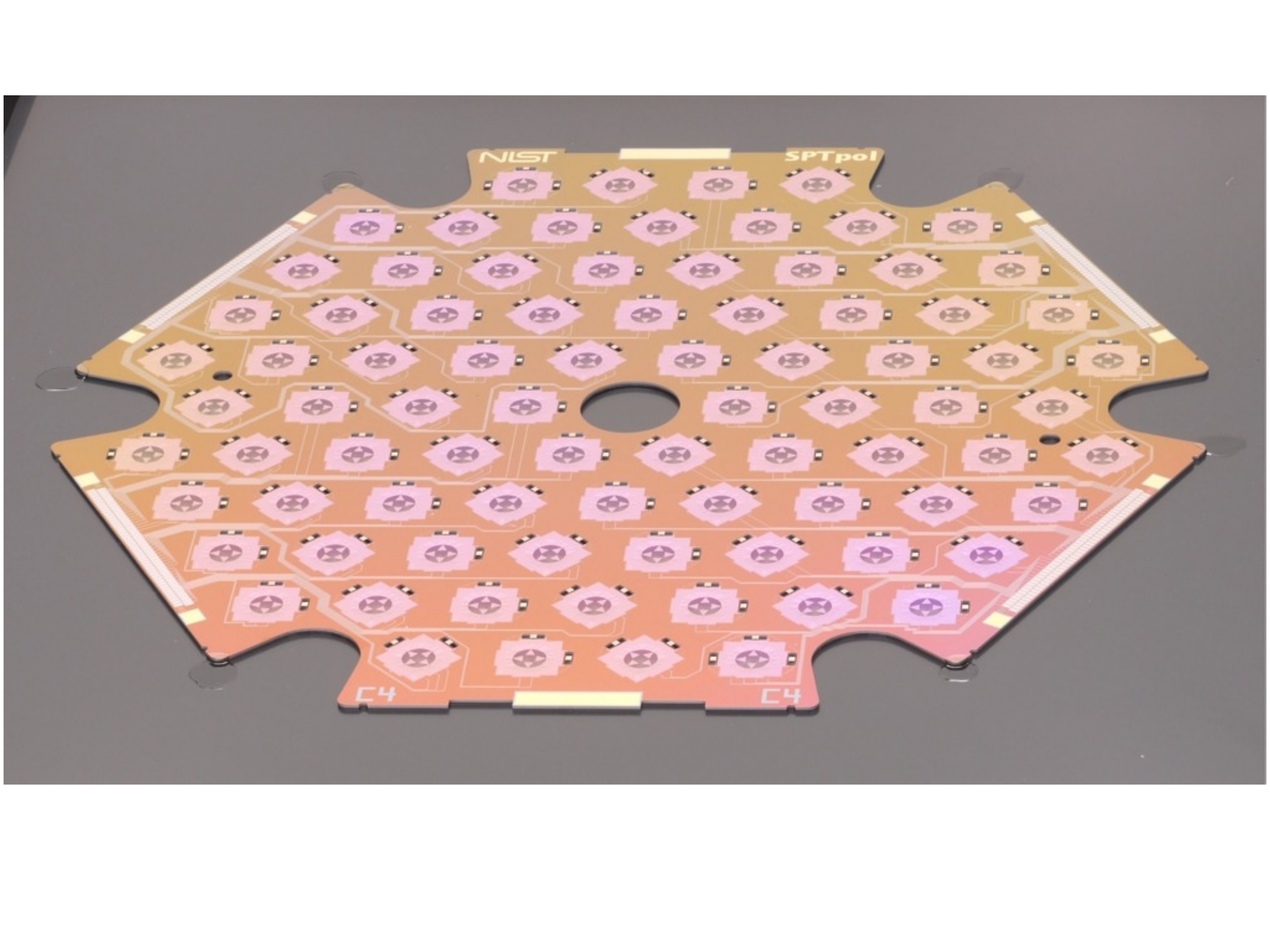}
\caption{An example of a microfabricated array of TES bolometers, an
  88 pixel, dual polarization TES bolometer array made at NIST for the
  SPTpol experiment. TES detectors are micro-machined from thin films
  deposited on silicon wafer substrates which means TES-based devices
  are fundamentally fabricated as arrays.}
\label{fig:tes_array}
\end{figure}

The TES has a number of strengths making it the best technology to
pursue for \cmbexp. First, TES detectors are fabricated via
micro-machining of thin films deposited on silicon wafer
substrates. As a consequence, the fundamental production unit for TES
devices are arrays of detectors (see Fig.~\ref{fig:tes_array}), an
important attribute when considering the production of the 500,000
detectors required by \cmbexp.  Second TES devices are low-impedance
($\le$1~$\Omega$) and can be multiplexed with modern-day
Superconducting QUantum Interference Device (SQUID)
multiplexers~\cite{fmux,tmux,umux}. Multiplexed readouts are important
for operating large detector arrays at sub-Kelvin temperatures and are
essential for \cmbexp. Lastly, TES detectors have been successfully
deployed as focal planes at the forefront of CMB measurements.

The TES was invented by HEP for detecting Dark Matter and
neutrinos. Its subsequent integration into CMB focal planes has
enabled kilo-pixel arrays realizing the Stage II CMB program and
ushering in an era of unprecedented sensitivity. TES-based CMB
detectors are the favored technology among Stage II and proposed Stage
III experiments, and have a clear path to the sensitivities required
by \cmbexp. The ubiquity of TES detectors for CMB illustrates the
direct connection between HEP-invented technology and CMB science.

\subsection*{The \cmbexp\ Experimental Program}
Delivering a half-million background-limited bolometers necessitates a
change in the execution of the US ground-based CMB program. The
current US program consists of a number of independent (primarily
university) efforts, each focused on the development and delivery of
their own instrument. The involvement of HEP in the current process
has been through small investments targeted at specific technical
contributions. Realizing \cmbexp\ will require a radical change in
this approach where HEP resources take the leading role. Of particular
importance is an increased participation and support of national labs
to provide resources which are unavailable to university groups. In
particular, 1) leveraging micro-fabrication tools and expertise
available at multi-purpose national labs is essential for the
successful fabrication of the \cmbexp\ detectors, and 2) supporting
the computing infrastructure to support the greatly increased data
rate, dataset size and analysis complexity.

The \cmbexp\ experimental program will build on the success of Stage
II and Stage III CMB experiments.  It will be a coherent effort
incorporating resources from both national labs and university groups.

The major instrumentation challenges are:

\begin{itemize}

\item {\bf Improved Production Reliability} The favored technology for
  \cmbexp\ are TES bolometers coupled through superconducting
  microstrip. Critical for the TES technology is reliable and optimal
  superconducting microstrip performance at millimeter
  wavelengths. Recent work on microstrip-coupled CMB detectors have
  demonstrated that it is possible to make superconducting microstrip
  which is virtually loss-less at the required
  frequencies~\cite{sptpol150}; however, the fabrication yield needs
  to be improved for \cmbexp. Thus, one of the principle components of
  the \cmbexp\ program is developing a reliable mass production
  process. Such work requires well maintained tooling, dedicated
  materials deposition, and understanding and control of all the
  materials dependent loss mechanisms.

\item {\bf Increased Production Volume and Throughput} Achieving
  $O$(500,000) TES detectors demands new investment into TES array
  production resources. The required production throughput needs
  access to micro-fabrication resources with exclusive control of the
  thin film deposition systems. This exclusive access to
  microfabrication tooling falls squarely within the domain of
  national labs. Additionally, an extensive program of detector
  testing, characterization, and quality control is crucial for the
  mass production of 500,000 TES bolometers. This requirement will be
  met by establishing test facilities and organizing a quality
  assurance program among the universities and national labs.

\item {\bf Multiplexed TES Readout} Multiplexed TES readouts are
  required for implementing focal planes with more than 1000 detector
  elements and will continue to be an active component of the
  \cmbexp\ R\&D program. Modest improvements over existing fielded
  Stage-II multiplexer technology will be sufficient for the needs of
  \cmbexp. However, recent developments with microwave-based readout
  techniques for TES detetors may lead to new multiplexer technologies
  with broader applicability and lower cost, and could be synergistic
  with mKID readout development efforts.

\item {\bf Large Cryogenic Optics} The large size of the
  \cmbexp\ focal planes together with the required sub-Kelvin
  operating temperature necessitates the development of new broad-band
  large aperture refractive optics which permit high throughput at
  millimeter wavelengths while blocking infrared thermal
  emission. Upcoming Stage-III experiments will serves as a proving
  ground for some new cryogenic optics technologies. \cmbexp\ will
  build on these Stage-III accomplishments with the goal of developing
  manufacturing techniques to yield a large number of customized
  cryogenic optics with optimal performance.

\end{itemize}

The \cmbexp\ program also presents equally challenging requirements in
computing infrastructure and analysis tools.  The analysis of a CMB
dataset typically proceeds in sequence of steps:
\begin{itemize} 
\item[-]{\bf Map-making} reduces the detectors' samples to maps of the
  temperature and polarization (typically split into $Q$ and $U$
  modes) of the observed sky,
\item[-] {\bf Component separation} extracts the CMB from any
  foreground contaminants in these maps,
\item[-] {\bf Power spectrum estimation} reduces the CMB maps to the
  six auto ($TT$, $EE$, $BB$) and cross ($TE$, $TB$, $EB$) angular
  power spectra of the temperature and polarization (now expressed as
  $E$ and $B$ modes),
\item[-] {\bf Parameter estimation} determines the likelihoods of the
  parameters of any chosen cosmological model given these power
  spectra.
\end{itemize}
This is essentially a series of data compressions, from the time- to
the pixel- to the multipole- to the parameter-domain, and the methods
we are able to employ to perform each of these steps will depend on
their computational costs.

The CMB-S4 experiment presented here will field $O$(500,000) detectors
each sampling at 100 Hz for 5 years, during which it will repeatedly
scan half of the sky at 3 arcminute resolution or better. Assuming a
70\% duty cycle, this will yield a dataset with ${\cal N}_t \sim 6
\times 10^{15}$ time samples over ${\cal N}_p \sim 6 \times 10^{8}$
sky pixels. Note that the science goals of this experiment require
10,000 times as many observations per pixel as the current Planck
satellite mission.

Under minimal assumptions, we can write simple Gaussian likelihoods
for the maps given the data and the noise correlations in the time
domain, and for the power spectra given the maps and the noise
correlations in the pixel domain. However, the number of operations
required first to construct the dense pixel domain noise correlation
matrix, and then to use it to maximize the spectral likelihood, both
scale as the cube of the number of pixels in the map; since ${\cal
  N}_{p}^{3} \sim 2 \times 10^{26}$ here, this approach is clearly
unrealistic.

The alternative is the pseudo-spectral approach, where the power
spectra are estimated from the maps ignoring their inhomogeneous
noise, incomplete sky coverage, and any filtering applied to remove
parasitic signals such as atmospheric emissions. The resulting
pseudo-spectra are then corrected for the biases this introduces by
applying a transfer function. Ideally this function is determined by
Monte Carlo methods, simulating many realizations of the data and
analyzing each in the same way as the real data. Comparing the
simulations' known input spectra and derived output pseudo-spectra we
can estimate the transfer function, which can then be inverted and
applied to the real pseudo-spectra. Dominated by the iterative
map-making -- whether destriping or maximum likelihood -- this
analysis scales linearly with the numbers of realizations, iterations
and samples, for an overall operation count of ${\cal N}_r \, {\cal
  N}_i \, {\cal N}_t \sim 6 \times 10^{21}$, assuming 10,000
realizations (for 1\% uncertainties) each requiring 100 iterations.

While this is a daunting number, it is consistent with the exponential
growth in the size of CMB datasets over the last 20 years (see
Snowmass Cosmological Computing report). Mirroring as it does Moore's
Law for computational capability, this growth constrains us to stay on
the leading edge of high performance computing if we wish to realize
the full scientific potential of the datasets we are
gathering. Staying on that leading edge -- wherever it may take us --
represents the core challenge of CMB data analysis. To date this has
necessitated following the evolution from simple increases in
processor clock speed, to higher and higher core counts, to
increasingly heterogenous accelerator-based systems. Beyond that, on
the timescale of a CMB-S4 mission, we can only speculate, although we
can be confident that exploiting the full capability of these systems
will become more and more challenging as they become increasingly
inhomogeneous, power-constrained, and data-bound, and will require
continued investment by HEP in the computational science underpinning
these analyses.

\subsection*{The role of the national laboratories in CMB experiment}
\cmbexp\ is the leading future CMB experiment. It is an ambitious
undertaking where the primary technical challenge is one of scale,
both in detectors and the resulting data set.  For reference, current
Stage-II instruments have focal planes with $O$(1000) elements whereas
\cmbexp\ will have $O$(500,000).  To address the technical challenges
of \cmbexp, HEP must re-envision its role in CMB experiments. Critical
for the success of \cmbexp\ is significant participation by national
labs. This increased participation will be complemented with an
organized program with university groups. The focus of the
\cmbexp\ technical R\&D will be the development of essential
technology for the mass production and operation of the 500,000 TES
detectors, and analysis of the resulting large data set.

\cmbexp\ will have a profound impact on our understanding of High
Energy Physics. It will provide the best path forward for
investigating inflation, i.e., through the measurement of $B$-mode
polarization of the CMB imprinted by inflationary gravitational waves,
while also using CMB lensing to lead to measurement of the sum of the
neutrino masses with an uncertainties of 16 meV. \cmbexp\ will be the
CMB experiment of the upcoming decade and it is critical for
understanding cosmology and high energy physics in the post-{\sl
  Planck} era.

\bibliography{bibCF5}{}

\end{document}